\documentclass[10pt,twocolumn,notitlepage,aps,pra,showpacs,superscriptaddress,longbibliography,nofootinbib]{revtex4-2} 

\usepackage[T1]{fontenc}% optional T1 font encoding

\usepackage[utf8]{inputenc} 
\usepackage{amssymb,mathtools}
\usepackage{graphicx}
\usepackage{amsthm}
\usepackage{amssymb, latexsym, bm, mathtools, braket, multirow,enumitem}

\usepackage[cmintegrals]{newtxmath}
\usepackage{color}

\theoremstyle{plain}
\newtheorem{lemm}{Lemma}

\newtheorem{theo}{Theorem}

\theoremstyle{definition}
\newtheorem{remark}{Remark}
\newtheorem{prot}{Protocol}

\def\FF{\mathbb{F}}
\def\ZZ{\mathbb{Z}}

\DeclareMathOperator{\Tr}{Tr}

\newcommand\sI{\mathsf{I}}

\newcommand\sR{\mathsf{R}}
\newcommand\sS{\mathsf{S}}

\newcommand\sV{\mathsf{V}}
\newcommand\sX{\mathsf{X}}
\newcommand\sY{\mathsf{Y}}
\newcommand\sZ{\mathsf{Z}}

\newcommand\sT{\mathsf{T}}
\newcommand\sH{\mathsf{H}}
\newcommand{\sCZ}{\mathsf{C_Z}}
\newcommand{\sCnot}{\mathsf{C_{not}}}
\newcommand\bX{\bm{X}}
\newcommand\bT{\bm{T}}
\newcommand\bH{\bm{H}}
\newcommand{\bCZ}{\bm{C_Z}}
\newcommand\bU{\bm{U}}
\newcommand\bW{\bm{W}}
\newcommand\bV{\bm{V}}
\newcommand\bQ{\bm{Q}}

\newcommand\bGamma{\bm{\Gamma}}

\newcommand\bx{\bm{x}}
\newcommand\by{\bm{y}}
\newcommand\bz{\bm{z}}
\newcommand\bw{\bm{w}}
\newcommand\ba{\bm{a}}
\newcommand\bb{\bm{b}}
\newcommand\bc{\bm{c}}
\newcommand\bk{\bm{k}}
\newcommand\be{\bm{e}}

\def\Label#1{\label{#1}\ [\ \text{#1}\ ]\ }
\def\Label{\label}

\begin{document}
\title{Oblivious Quantum Computation and Delegated Multiparty Quantum Computation}

\author{Masahito Hayashi}
\email{hmasahito@cuhk.edu.cn, hayashi@iqasz.cn}
\affiliation{School of Data Science, The Chinese University of Hong Kong, Shenzhen, Longgang District, Shenzhen, 518172, China}
\affiliation{International Quantum Academy (SIQA), Futian District, Shenzhen 518048, China}
\affiliation{Graduate School of Mathematics, Nagoya University, Chikusa-ku, Nagoya, 464-8602, Japan}

\begin{abstract}
We propose a new concept, oblivious quantum computation, which requires performing oblivious transfer with respect to the computation outcome of the quantum computation, where
the secrecy of the input qubits and the program to identify the quantum gates are required.
Exploiting quantum teleportation,
we propose a two-server protocol for this task, which realizes an exponential improvement
for the communication complexity over the simple application of two-server (quantum)
oblivious transfer to the sending of the computation result.
Also, we discuss delegated multiparty quantum computation,
in which, several users ask multiparty quantum computation to server(s)
only using classical communications.
We propose a two-server protocol for the latter task as well.
\end{abstract}

\maketitle

\section{Introduction}
Recently, quantum computation attracts much attention from various research areas.
However, exponential improvement by quantum computer is very limited %limited to several computation tasks 
\cite{Shor}.
This paper focuses on oblivious transfer for computation outcome 
and derives an exponential improvement for this task
because
oblivious transfer is a cryptographic primitive task.
Consider that the servers have a secret computer program $\bw$, 
and the user wants its computation outcome $f_{\bw}(\bk)$ only with a specific input $\bk
\in \ZZ_2^n$,
where $f_{\bw} $ is the function based on the computer program $\bw$.
In this case, the user wants to hide the input $\bk\in \ZZ_2^n$,
and the servers want to hide the program $\bw$, i.e., 
want to keep its information except for the computation outcome $ f_{\bw}(\bk)$. 
 
If they employ oblivious transfer \cite{Rabin}, this task can be realized
while oblivious transfer is often called symmetric information retrieval \cite{GIKM}.
Although one-server oblivious transfer is impossible information-theoretically \cite{Mayers,Lo},
it is information-theoretically possible with two servers
if they share random numbers or entangled states.
However, even when we employ quantum communication,
information-theoretically secure oblivious transfer requires
linear communication size with respect to the number of possible inputs \cite{KdW03, KdW04}.
%That is, if the input is given as $n$ bits, it requires communication complexity $O(2^n)$.
Although the papers \cite{SH19} proposed its efficient quantum protocol,
it works well when the number of possible choices by the user is fixed and only the size of each 
message increases. 
That is, the protocol \cite{SH19} does not work in this case.
When we apply this method, the required communication complexity is
linear in $2^n$ because the input is composed $n$ bits.
In addition, in this scenario, the server(s) need to derive the computation outcomes 
for all $2^n$ inputs, which requires exponential computation time.

In this paper, to realize its exponential improvement, 
we propose a new conceptual task, oblivious quantum computation (OQC),
and a concrete protocol to realize this task with two servers, which is called a two-server OQC (TOQC) protocol.
Exploiting quantum teleportation \cite{tele},
our TOQC protocol realizes polynomial complexity by using quantum computation and quantum communication
when 
the number of servers is two and 
the computation requires polynomial time complexity and polynomial space complexity
with respect to the input size $n$.
Our protocol can be considered as a quantum computation version of oblivious transfer as follows.

Consider that the servers have a secret program $\bw$ to realize unitary operation $W(\bw)$
on $n$ qubits,
and the user wants 
a specific subsystem of the output state $W(\bw)|\psi\rangle$ only with a specific input state 
$|\psi\rangle \in {\cal H}^{\otimes n}$, where
${\cal H}$ is one qubit system $\mathbb{C}^{2}$.
Then, the user wants to hide the input state $|\psi\rangle \in {\cal H}^{\otimes n}$,
and the servers want to hide the program $\bw$, i.e., 
want to keep its information except for the output state $W(\bw)|\psi\rangle$.
We call the above task oblivious quantum computation.
This task can be considered as a kind of secure quantum computation.
When our unitary $W(\bw)$ is given as a combination of 
the controlled Z operation, a modification of the Hadamard gate, and the $1/8$-phase gate $\sT$
on the input qubits system ${\cal H}^{\otimes n} $,
which forms a universal gate set,
we propose an efficient protocol for this task with two servers.
When the number of quantum gates is $m$,
several prior entangled states are prepared between the two servers,
and both servers are not allowed to communicate with each other,
the communication complexity of our protocol is upper bounded by 
$(2n^2+ 20 n) m$ bits plus $2n$ qubits.
Hence, when the number of quantum circuits is polynomial for the input qubit length $n$,
the communication complexity of our protocol is polynomial. 
When the classical computation $f_{\bw}(\bk)$ can be realized by quantum computation with 
polynomial computation space and polynomial number of quantum gates,
our protocol offers an exponential improvement 
over simple application of oblivious transfer with multiple servers \cite{KdW03, KdW04,SH19,SJ17,SJ17-2}.
In fact, since our protocol has no information leakage, it 
has stronger secrecy than cheat-sensitive secrecy given by \cite{GLM}.

In addition, we discuss delegated multiparty quantum computation,
in which, several users ask multiparty quantum computation to server(s)
only by using classical communications.
The conventional delegated quantum computation \cite{Childs,BFK,BKBF,MF,Morimae,MDF,MF2,LCWW,SZ,HM}
realizes single-user computation with the blindness condition.
The delegated multiparty quantum computation requires the servers to realize
secure multiparty computation \cite{CGS,BCGHS,DGJMS}
with the blindness condition.
In this problem setting, it is required that a user cannot obtain any information for other users
except for the computation outcome.
Although the paper \cite{KKMO} proposed a similar problem,
it allows the users to use single qubit operations and quantum communications.
Further, combining oblivious quantum computation
and delegated multiparty quantum computation, we propose another new concept,
generalized delegated multiparty quantum computation as a unified concept.
This concept contains delegated multiparty quantum computation and a variant without quantum communication of 
oblivious quantum computation as special cases.
Since one-server delegated quantum computation is impossible \cite{DK,MK,MNTT},
we propose a two-server protocol for generalized delegated multiparty quantum computation
by modifying our two-server protocol for oblivious quantum computation.

The remainder of this paper is organized as follows.
To explain the key idea of our protocol,
Section \ref{S2} introduces a toy protocol, which 
explains how quantum teleportation works for our aim. 
This idea takes a key role in our protocols.
Section \ref{S3} prepares various notations used in this paper.
Using these notations,
Section \ref{S4} introduces several concepts including 
the definitions of the tasks of oblivious quantum computation,
delegated multiparty quantum computation,
and generalized delegated multiparty quantum computation.
Section \ref{S7} introduces our protocol for oblivious quantum computation, and shows its correctness, its user-secrecy, and its server-secrecy.
 Section \ref{S8-2} describes our protocol for 
generalized delegated multiparty quantum computation,
and shows
its correctness, its user-secrecy, and its server-secrecy.
Finally, Section \ref{S9} gives the conclusion.

\section{Toy protocol}\Label{S2}
To explain the basic idea of this paper,
we consider a toy protocol, where 
the random Pauli operation works as masking the input quantum state,
and quantum teleportation works as state transfer.
For this aim, we define 
the flip operator $\sX$,
the phase gate $\sZ$,
and $1/8$-phase gate $\sT$ as
\begin{align}
\sX:=
\left(
\begin{array}{cc}
0 & 1\\
1& 0
\end{array}
\right) ,\quad
\sZ:=
\left(
\begin{array}{cc}
1 & 0\\
0& -1
\end{array}
\right), \quad
\sT:=
\left(
\begin{array}{cc}
1 & 0\\
0& e^{i \pi/4}
\end{array}
\right).
\end{align}
Then,
we consider the case when the secret unitary is given as 
$\sT^y$ on a single qubit with $y \in \ZZ_8$.
The user wants the output state $\sT^y |\psi\rangle$ only with a specific input state 
$|\psi\rangle \in {\cal H}$, but 
the user wants to hide the input state $|\psi\rangle \in {\cal H}$.
In contrast, the servers want to hide the information $y \in \ZZ_8$, i.e., 
the user obtains no other information for the output state $\sT^y |\psi\rangle $.
This task can be realized as follows when both servers are not allowed to communicate with each other.
\if0
We define the controlled-not unitary gate $\sV$ on ${\cal H}^{\otimes 2}$ as
\begin{align}
\sV|j\rangle |j'\rangle =|j\rangle |j'+j\rangle,
\end{align}
which implies the relation $\sV|j\rangle |0\rangle=|j\rangle |j\rangle$.
When its domain is restricted to the subspace spanned by $|00\rangle|10\rangle $,
the operator is an isometry and is denoted by $\hat{\sV}$.
\fi

\begin{prot}[Toy protocol] \Label{toy}
The following protocol realizes the above task.
\begin{description}[leftmargin=1.5em]
\item[0)] \textbf{Preparation}: 
Servers A and B share the entangled state 
$|\Phi\rangle =
\frac{1}{\sqrt{2}}(|0\rangle|0\rangle 
+|1\rangle|1\rangle ) \in {\cal H}_A\otimes {\cal H}_B$.

\item[1)] \textbf{Query 1}: 
The user generates a quantum state $|\psi\rangle \in {\cal H}$
and random bits $A_0,B_0\in\ZZ_2$ 
and random variables $Q_u \in\ZZ_8$ for $u \in \ZZ_2$
according to the uniform distribution. 
The user applies $\sZ^{B_0} \sX^{A_0}$, and obtains
the state $\sZ^{B_0}\sX^{A_0} |\psi\rangle \in {\cal H} $.
Then, the user sends $Q_u$ and 
the system ${\cal H} $ to Server A.

\item[2)] \textbf{Answer 1}: 
Server A applies the unitary $ \sT^{Q_0 y} (\sX \sT^{Q_1 y} \sX)$
to ${\cal H}$.
Server A applies the Bell measurement 
$\{|\Phi_{a,b}\rangle\}_{a,b \in \ZZ_2}$
on ${\cal H}\otimes {\cal H}_A$,
where $|\Phi_{a,b}\rangle:=
(\sX^a \sZ^b) \otimes I |\Phi\rangle$.
Then, Server A obtains the outcome $(A_1,B_1)$ and sends it to the user.

\item[3)] \textbf{Query 2}: 
The user generates random bits $Q_u' \in\ZZ_8$ as
\begin{align}
Q_{A_0+A_1}':= -Q_{A_0}+ 1 , \quad
Q_{A_0+A_1+1}':= -Q_{A_0+1},
\Label{AMMY}
\end{align}
and sends them to Server B.

\item[4)] \textbf{Answer 2}: 
Server B applies the unitary $ \sT^{Q_0' y} (\sX \sT^{Q_1' y} \sX)$
to ${\cal H}_B$ and sends ${\cal H}_B$ to the user.

\item[5)] \textbf{Construction}: 
The user applies $ \sZ^{B_0+B_1}\sX^{A_0+A_1} $ to the received state. 
\end{description}
\end{prot}

The correctness of the above protocol can be shown as follow.
Assume that the user and both servers are honest.
Then, the final state is calculated as 
%The state at the end of Step 2) is 
\begin{align}
&\sZ^{B_0+B_1}\sX^{A_0+A_1} 
\sT^{Q_0' y} (\sX \sT^{Q_1' y} \sX)\sX^{A_1} \sZ^{B_1} \nonumber \\
&\cdot \sT^{Q_0 y} (\sX \sT^{Q_1 y} \sX)\sZ^{B_0}\sX^{A_0} |\psi\rangle \nonumber \\
\doteq&
\sX^{A_0+A_1} 
\sT^{Q_0' y} (\sX \sT^{Q_1' y} \sX)\sX^{A_1} 
 \sT^{Q_0 y} (\sX \sT^{Q_1 y} \sX)\sX^{A_0} |\psi\rangle \nonumber \\
= &
(\sX^{A_0+A_1} \sT^{Q_0' y} \sX^{A_1+A_0})
(\sX^{1+A_1+A_0} \sT^{Q_1' y} \sX^{1+A_1+A_0}) \nonumber \\
&\cdot(\sX^{A_0} \sT^{Q_0 y} \sX^{A_0})(\sX^{1+A_0} \sT^{Q_1 y} \sX^{1+A_0}) |\psi\rangle \nonumber \\
= &
\sT^{Q_{A_0+A_1}' y} 
(\sX \sT^{Q_{1+A_1+A_0}' y} \sX )
\cdot \sT^{Q_{A_0} y} (\sX \sT^{Q_{1+A_0} y} \sX) |\psi\rangle \nonumber \\
\stackrel{(a)}{=} &
\sT^{ y}  |\psi\rangle ,
\end{align}
where $(a)$ follows from \eqref{AMMY}.
Here, $\doteq$ means the equal with a certain phase factor.

\if0
Since $\sX^u\sT^{Q_u y}\sX^u$ cancels $\sX^u\sT^{Q_u' y}\sX^u$ 
for $u\neq a$,
the resultant state after Step 2) is 
$(\sX^a \otimes \sX^a )
(\sT^{y} \otimes I)
\sV|\psi\rangle |0\rangle 
=(\sX^a \otimes \sX^a )
\sV (\sT^{y}|\psi\rangle) |0\rangle 
\in {\cal H}_A \otimes {\cal H}_B$.
The state after step 3) is 
$\sV^\dagger (\sX^a \otimes \sX^a ) (\sX^a \otimes \sX^a )
\sV (\sT^{y}|\psi\rangle)|0\rangle
=(\sT^{y}|\psi\rangle)|0\rangle$, which is the desired output state.
In this way, the protocol outputs the desired state when the user and both servers are honest.
\fi

Assume that the user is honest, and 
Servers A and B do not communicate with each other.
Server A receives only the system ${\cal H}$ and the variables $Q_0,Q_1$.
The variables $Q_0,Q_1$ are subject to the uniform distribution independently of 
$|\psi\rangle$ and $A_0,B_0$.
Hence, $Q_0,Q_1$ have no information for $|\psi\rangle$ and
$A_0,B_0$.
The average of the state 
$ \sZ^{B_0}\sX^{A_0} |\psi\rangle\langle \psi|(\sZ^{B_0}\sX^{A_0})^\dagger$
is the completely mixed state.
Hence, Server A has no information for $|\psi\rangle$.
Server B receives only the variables $Q_0',Q_1'$, which
are subject to the uniform distribution independently of $|\psi\rangle$.
Server B has no information for $|\psi\rangle$.

On the other hand, 
when the servers are honest,
the user obtains only one qubit system.
Hence, if the user obtains the desired state
$\sT^{ y}  |\psi\rangle $, the user cannot obtain any other information for $\sT$.

\section{Notation}\Label{S3}
To explain our problem setting of oblivious quantum computation,
we prepare several notations.
First, we define 
the modified Hadamard gate $\sH$
on a qubit system ${\cal H}$
and the operator $\sY$ as
\begin{align}
\sY:=\sZ\sX,\quad 
\sH:=
\frac{1}{\sqrt{2}}\left(
\begin{array}{cc}
1 & 1\\
-1& 1
\end{array}
\right).
\end{align}
Then, we have the commutation relation $\sY\sH=\sH\sY$.
The Hadamard gate is given as $\sT^4 \sH$.

Then, we define the controlled $Z$ gate $\sCZ$ 
on a two-qubit system ${\cal H}^{\otimes 2}$ as
\begin{align}
\sCZ:= \sum_{j,k\in \FF_2^2} (-1)^{jk}|j,k\rangle \langle j,k|.
\end{align}

It is known that 
the combination of
the controlled NOT gate, the phase gate $\sZ$, 
$1/8$-phase gate $\sT$, and the Hadamard gate $\sZ\sH$
forms a universal gate set \cite[Section 4.5.3]{NC}.
Since the controlled NOT gate is given as
a combination of $\sCZ,\sZ,\sH$
%$\sZ\sH \sCZ \sZ\sH$ 
and the phase gate $\sZ$ is given as $\sT^4$,
the set $\{\sH,\sT,\sCZ\}$ forms another universal gate set.
To see this fact, we consider operators acting on ${\cal K}:=\otimes_{s=1}^{n} {\cal H}_s$,
where the $s$-th system is written as ${\cal H}_s$.
%Next, we discuss the operators acting on ${\cal K}:=\otimes_{s=1}^{n} {\cal H}_s$,
%where the $s$-th system is written as ${\cal H}_s$.
When the operator $\sX$ acts on ${\cal H}_s$, it is written as $\sX_s$.
This rule is applied to other operators. 
In addition, the operator $\sCZ$ acts on ${\cal H}_s\otimes {\cal H}_t$,
it is written as $\sCZ_{(s,t)}$.
The controlled NOT gate $\sCnot_{(s,t)}$ on ${\cal H}_s\otimes {\cal H}_t$
is given as
$|0\rangle\langle 0|_s \otimes \sI_t+
|1\rangle\langle 1|_s \otimes \sX_t$.
Since $\sX=\sZ\sH \sH=\sZ\sH \sZ \sZ\sH$
and $\sI=\sZ\sH \sZ \sH$,
\begin{align}
\sCnot_{(s,t)}=&
\sZ_t\sH_t (|0\rangle\langle 0|_s \otimes \sI_t+
|1\rangle\langle 1|_s \otimes \sZ_t)\sZ_t\sH_t \nonumber\\
=&\sZ_t\sH_t \sCZ_{(s,t)} \sZ_t\sH_t
=\sT_t^4\sH_t \sCZ_{(s,t)} \sT_t^4\sH_t.
\end{align}
Also, since $-\sX=\sH \sH \sZ=\sH \sZ \sZ\sH \sZ$
and $-\sI=\sH \sZ \sH\sZ$, we have
\begin{align}
-\sCnot_{(s,t)}=&
\sH_t (|0\rangle\langle 0|_s \otimes \sI_t+
|1\rangle\langle 1|_s \otimes \sZ_t)\sZ_t\sH_t \sZ_t\nonumber\\
=&
\sH_t \sCZ_{(s,t)} \sT_t^4\sH_t \sT_t^4
\Label{ZKO}.
\end{align}
In this way, 
the set $\{\sH,\sT,\sCZ\}$ forms another universal gate set.

To handle these operators,
we define the sets $[n] = \{1,\ldots, n\}$ and $[n]_2 = \{(s,t)\}_{ s<t \in [n]}$.
\if0
We define 
$\bV:=\prod_{s \in [n]}\sV_{s}$ and 
$\hat{\bV}:=\prod_{s \in [n]}\hat{\sV}_{s}$.
\fi
For $\bx=(x_s)_{s \in [n]} \in\ZZ_8^n$ and $\by=(y_s)_{s \in [n]} \in\ZZ_8^n$, we define
\begin{align}
\bH(\bx):=\prod_{s \in [n]}\sH_s(x_s),\quad
\bT(\by):=\prod_{s \in [n]}\sT_s(y_s).
\end{align}
Here, $\sH_s(x)$ expresses $\sH_s^x$, and this rule is applied to other operators.
This rule is useful when $x$ has a complicated form including various indexes.

We define the set $\ZZ_8^{[n]_2}:=\{ ( z_{(s,t)}  )_{(s,t) \in [n]_2}\}$, where
$z_{(s,t)} \in \ZZ_8 $.
For $\bz\in \ZZ_8^{[n]_2}$, we define
\begin{align}
\bCZ(\bz):=\prod_{(s,t) \in [n]_2}\sCZ_{(s,t)}(z_{(s,t)}).
\end{align}
In particular, we define
\begin{align}
\bU(\bx,\by,\bz):=
\bH(\bx)
\bT(\by)
\bCZ(\bz).
\end{align}

Since the gate set $\{\sH,\sT,\sCZ\}$
is a universal gate set, 
universal quantum computation can be performed by 
unitary operation
\begin{align}
\bW(\bw)
%:=\prod_{j=1}^{m}\bU(\bx_j,\by_j,\bz_j).
%\prod_{j=1}^{m}\bU(\bx_j,\by_j,\bz_j)
:=\bU(\bx_m,\by_m,\bz_m)\cdot \ldots\cdot\bU(\bx_1,\by_1,\bz_1),
\end{align}
where
$\bw:=
((\bx_m,\by_m,\bz_m), \ldots,
(\bx_1,\by_1,\bz_1))$. 
Hence, the string 
$\bw$
can be considered as a program for quantum computation
because it identifies the unitary operation.

Given two programs $\bw, \bw'$,
we define the product $\bw\cdot \bw'$
as 
$((\bx_m\cdot \bx_m',\by_m\cdot \by_m',\bz_m\cdot \bz_m'), \ldots,
(\bx_1\cdot\bx_1',\by_1\cdot\by_1',\bz_1\cdot\bz_1'))$,
where 
$\bx_j\cdot \bx_j'$,
$\by_j\cdot \by_j'$,
$\bz_j\cdot \bz_j'$ are defined as
$(x_s x_s')_{s \in [n]} $,
$(y_s y_s')_{s \in [n]} $,
$(z_{(s,t)} z_{(s,t)}')_{(s,t) \in [n]_2} $.
When we define 
$\be:=
(\underbrace{(1,1,1), \ldots,(1,1,1)}_{m})$,
we have $\be\cdot \bw'=\bw'$.

When $m$ is written as $m_1+m_2$,
$\bw$ is divided into two parts. That is,
$\bw$ is written as $(\bw_1,\bw_2)$, 
where
$\bw_1:=((\bx_{m_1},\by_{m_1},\bz_{m_1}), \ldots, (\bx_1,\by_1,\bz_1))$
and
$\bw_2:=
((\bx_m,\by_m,\bz_m), \ldots, (\bx_{m_1+1},\by_{m_1+1},\bz_{m_1+1}))$. 
Then, we define
\begin{align}
\bW_1(\bw_1)
&:=\bU(\bx_{m_1},\by_{m_1},\bz_{m_1})\cdot \ldots\cdot\bU(\bx_1,\by_1,\bz_1) \\
\bW_2(\bw_2)
&:=\bU(\bx_{m},\by_{m},\bz_{m})\cdot \ldots\cdot\bU(\bx_{m_1+1},\by_{m_1+1},\bz_{m_1+1}) .
\end{align}

\section{Formulations}\Label{S4}
\subsection{Formulation of oblivious quantum computation}\Label{S41}
The problem setting of 
oblivious quantum computation 
is formulated as Fig. \ref{F1} with 
the following definitions of various concepts.

\subsubsection{Task}
Server(s) have a program $ 
\bw
:=((\bx_{m},\by_{m},\bz_{m}), \ldots, (\bx_{1},\by_{1},\bz_{1}))\in
(\ZZ_4^n \times \ZZ_8^n \times \ZZ_2^{[n]_2} )^{m}$.
When several server(s) exist, they are not allowed to communicate with each other.
While the user does not know it, 
the server(s) are required to apply 
$\bW(\bw)$ to the state $|\psi\rangle \in {\cal K}$.
The user wants 
the first $n_{\circ}$ qubits of
the output state $\bW(\bw)|\psi\rangle \in {\cal K}$.
We denote 
the first $n_{\circ}$ qubits of ${\cal K}$ by ${\cal K}_{\circ}$
and 
the remaining $n-n_{\circ}$ qubits of ${\cal K}$ by ${\cal K}_\triangle$.
Hence, the user wants 
the state $\Phi_{ideal}(\bw,|\psi\rangle):=
 \Tr_{{\cal K}_\triangle} \bW(\bw)|\psi\rangle\langle \psi|\bW(\bw)^\dagger$
on ${\cal K}_{\circ}$.
In the following, we use the subscript $\circ$ to identify the first $n_\circ$ systems.
%For example, we denote $\otimes_{s=1}^{n_\circ} \sV_s$ by $\bV_{\circ}$.

\if0
\begin{figure}[tb]{}
    \begin{tabular}{cc}
      \begin{minipage}[t]{0.3\hsize}
        \centering
        \includegraphics[keepaspectratio, scale=0.5]{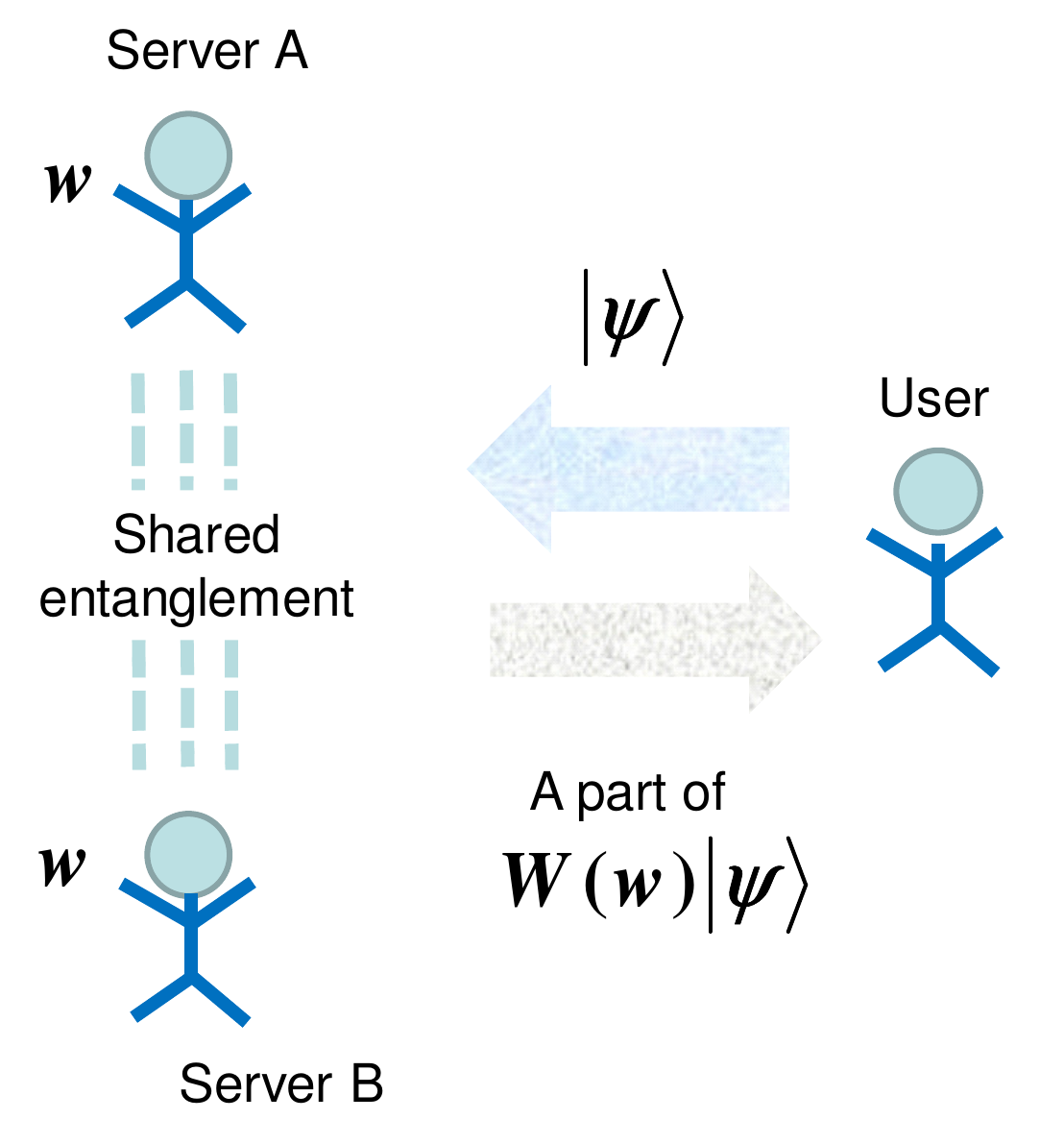}
\caption{{\bf Oblivious quantum computation with two servers.}
This figure shows a protocol for oblivious quantum computation
with two servers when the two servers share an entangled state.
}\Label{F1} 
      \end{minipage} &
      \begin{minipage}[t]{0.32\hsize}
        \centering
        \includegraphics[keepaspectratio, scale=0.5]{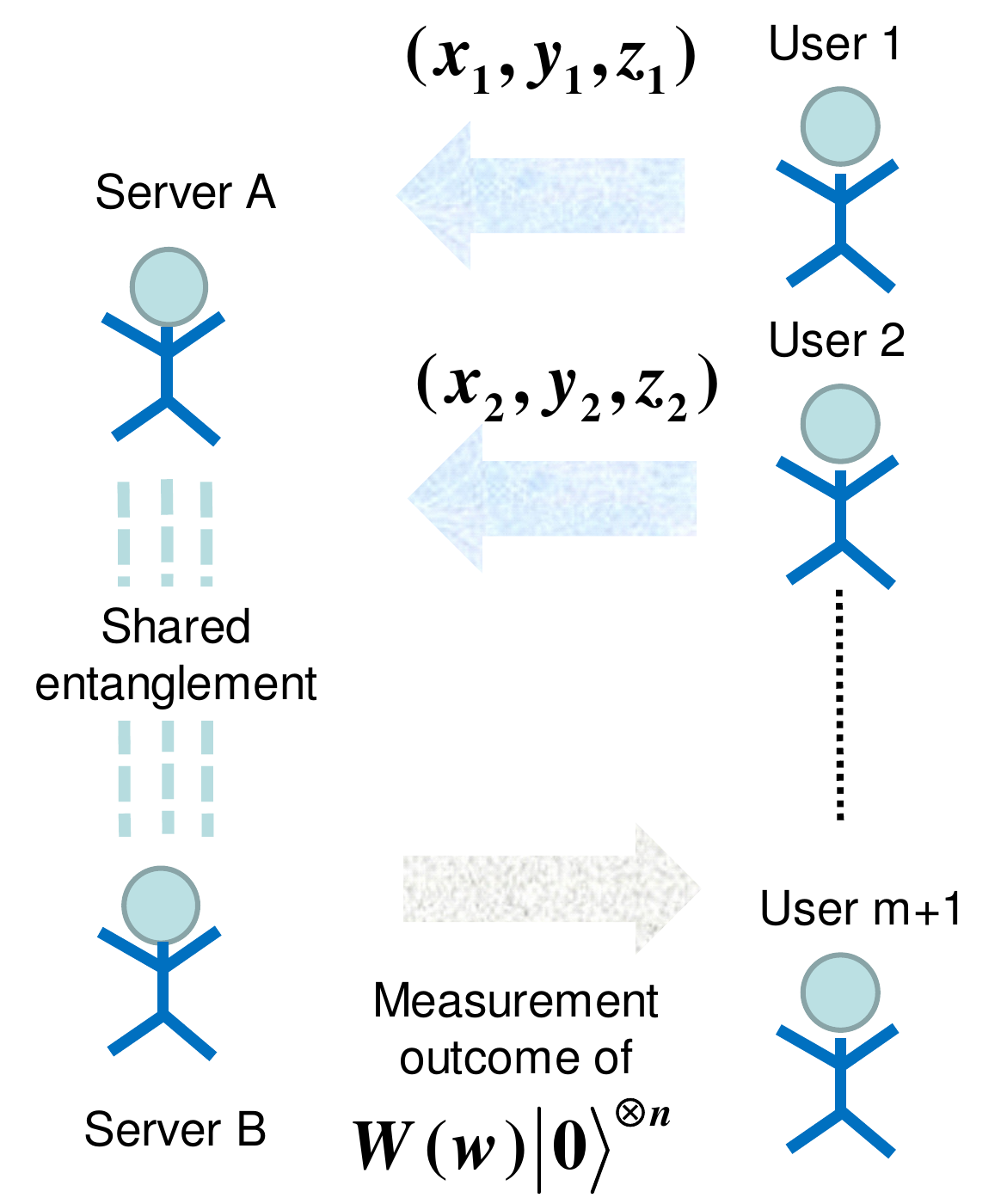}
\caption{{\bf Delegated multiparty quantum computation with two servers.}
This figure shows a protocol for delegated multiparty quantum computation
with two servers when the two servers share an entangled state.
}\Label{F2} 
      \end{minipage}
      \begin{minipage}[t]{0.32\hsize}
        \centering
        \includegraphics[keepaspectratio, scale=0.5]{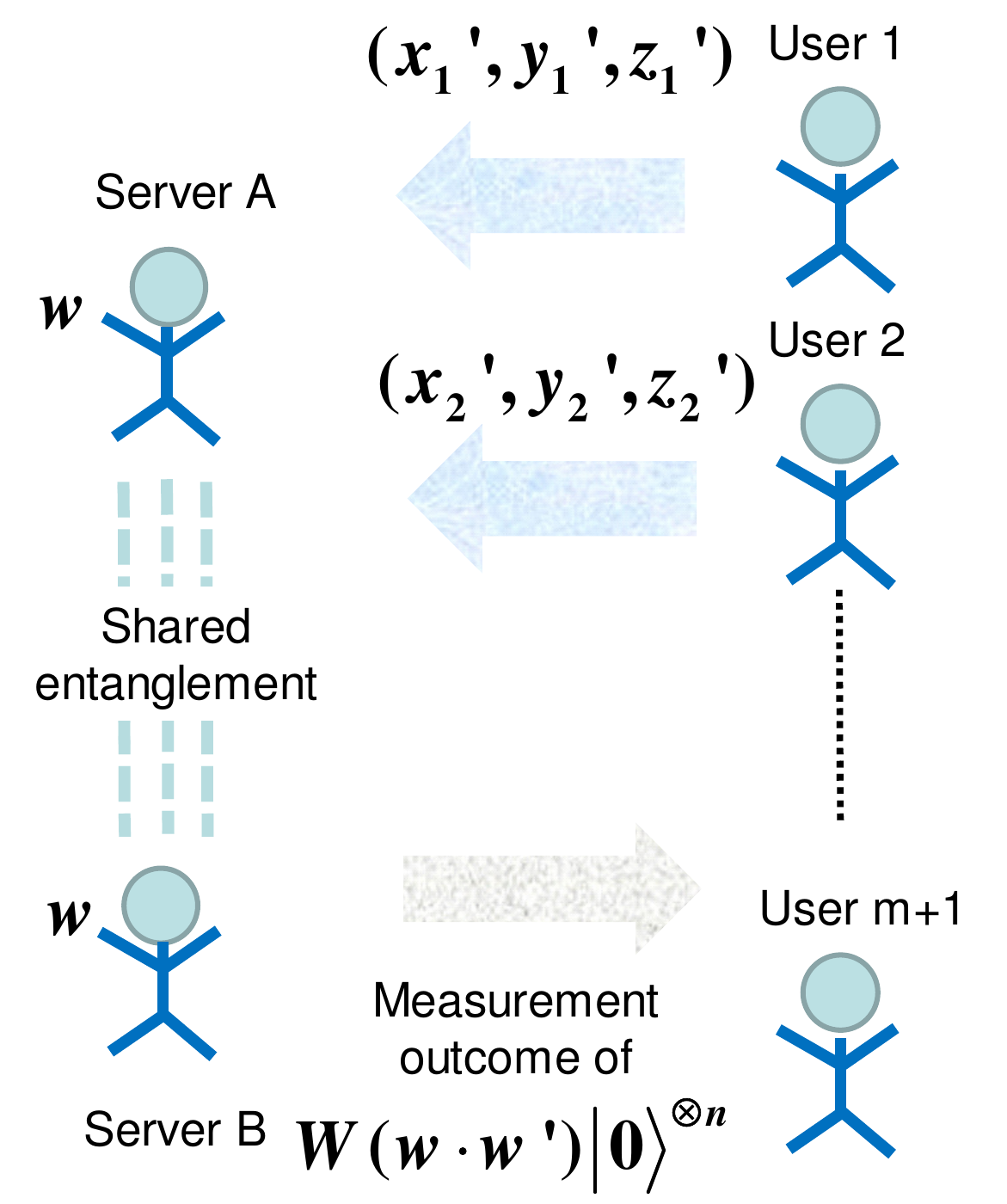}
\caption{{\bf Generalized delegated multiparty quantum computation with two servers.}
This figure shows a protocol for generalized delegated multiparty quantum computation
with two servers when the two servers share an entangled state.
}\Label{F3} 
      \end{minipage}
    \end{tabular}
  \end{figure}
\fi

\begin{figure}[tb]{}
\begin{center} 
\includegraphics[width=\linewidth]{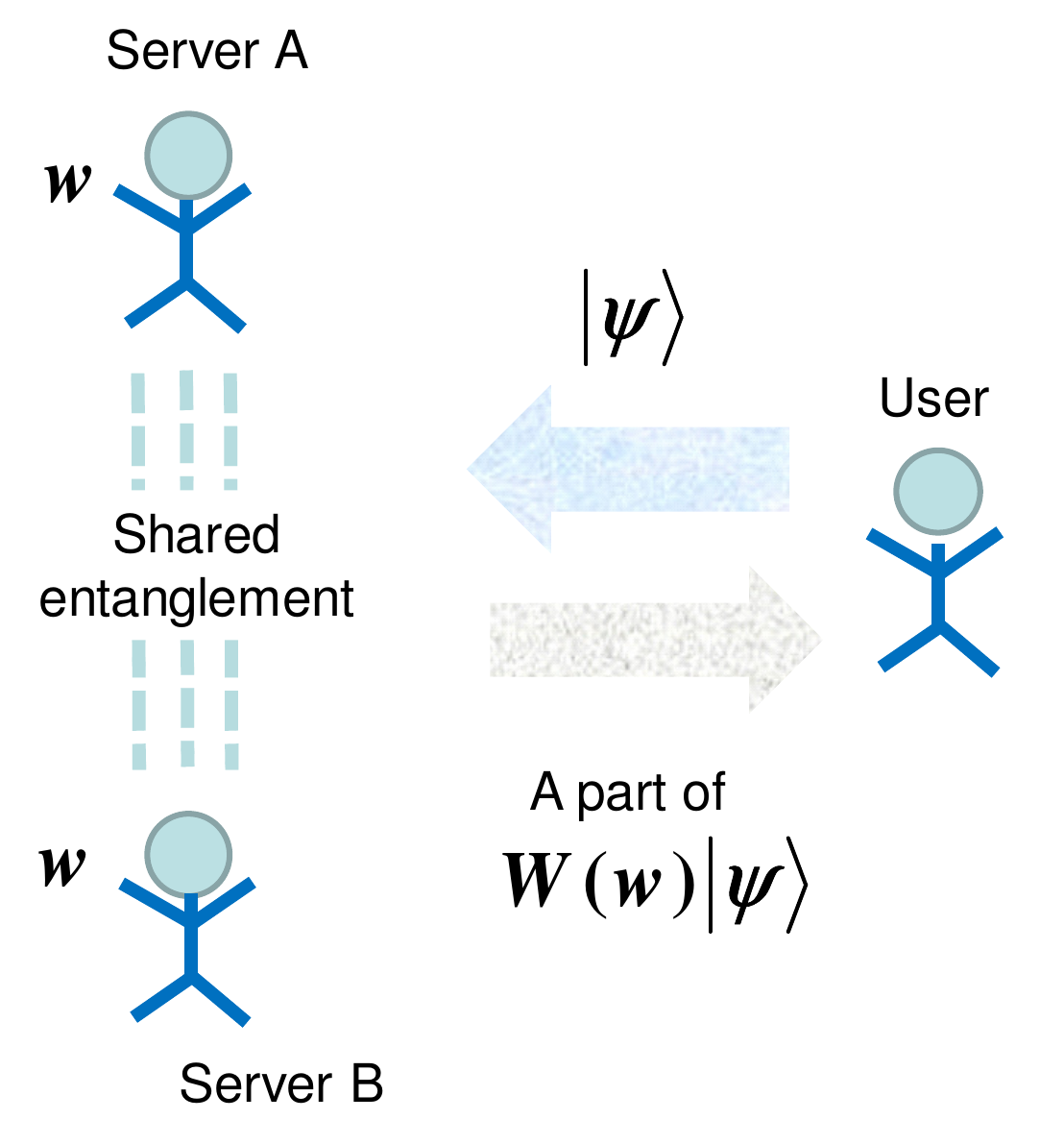}
\end{center}
\caption{{\bf Oblivious quantum computation with two servers.}
This figure shows a protocol for oblivious quantum computation
with two servers when the two servers share an entangled state.
}\Label{F1} 
\end{figure}

\subsubsection{Correctness and complexity}
An OQC  protocol $\Phi$ has two types of inputs.
The first input is the program 
$\bw \in 
(\ZZ_4^n \times \ZZ_8^n \times \ZZ_2^{[n]_2} )^{m}$.
The second input is the input quantum state $|\psi\rangle \in {\cal K}$.
The output of the protocol is a state $\rho_{out}$ on ${\cal K}$,
which is desired to be $\Phi_{ideal}(\bw,|\psi\rangle)$.

An OQC  protocol $\Phi$ has bilateral communication.
The upload communication is the communication from the user to the servers,
and  
the download communication is the communication from the servers to the user.

The communication complexity is composed of 
the upload complexity and the download complexity.
The upload complexity is the sum of the communication sizes of all upload communications, and
the download complexity is the sum of the communication sizes of all 
download communications.
The sum of the upload and download complexity is called
the communication complexity.
For a OQC protocol $\Phi$, we denote the output state 
by $\Phi_{out}(\bw,|\psi\rangle)= \rho_{out}$.
The upload complexity, the download complexity, and the communication complexity are 
denoted by $UC(\Phi)$, $DC(\Phi)$, and $CC(\Phi)$, respectively.
Hence, the communication complexity $CC(\Phi)$ is calculated as
$UC(\Phi)+DC(\Phi)$.

An OQC protocol $\Phi$ is called correct when 
%the following condition holds. When the user and the servers are honest,
the relation $\Phi_{out}(\bw,|\psi\rangle)=
\Phi_{ideal}(\bw,|\psi\rangle)
%\Tr_{{\cal K}^2} \bW(\bw)|\psi\rangle\langle \psi|\bW(\bw)^\dagger
$ holds
for any state $|\psi\rangle \in {\cal K}$ and
$\bw\in 
(\ZZ_4^n \times \ZZ_8^n \times \ZZ_2^{[n]_2} )^{m}$.

\subsubsection{Two types of secrecy}
An OQC  protocol $\Phi$ has two types of secrecy.
One is the user-secrecy, and the other is the server-secrecy.
We say that 
an OQC  protocol $\Phi$ satisfies the user-secrecy 
when the following condition holds. 
To explain the user-secrecy, for $J=A,B$,
we denote the final state on Server $J$ dependently of the input state $|\psi\rangle$
by $\rho_{Y_J||\psi\rangle}$. 
When the user is honest,
no server obtains the information of the user's input state $|\psi\rangle$, i.e., 
the relation 
\begin{align}
\rho_{Y_J||\psi\rangle}
=
\rho_{Y_J||\psi'\rangle}
\label{MLK}
\end{align}
holds for $J=A,B$ and 
any states $|\psi\rangle,|\psi'\rangle \in {\cal K}$.

We say that an OQC  protocol $\Phi$ satisfies the server-secrecy 
when the following condition holds. 
When the servers are honest and the output state $\rho_{out}$ equals 
$\Phi_{ideal}(\bw,|\psi\rangle)$,
the user obtains no information for 
the program
$\bw \in 
(\ZZ_4^n \times \ZZ_8^n \ZZ_2^{[n]_2} )^{m}$
except for the desired output state 
$\Phi_{ideal}(\bw,|\psi\rangle)$, i.e.,
the user can generate the final state on his/her own whole system
by using $\Phi_{ideal}(\bw,|\psi\rangle)$,
the classical information describing the initial state $|\psi\rangle$,
and classical information generated by himself/herself.

\subsection{Formulation of delegated multiparty quantum computation}\Label{S42}
The problem setting of delegated multiparty quantum computation
(DMQC) is formulated as Fig. \ref{F2}
with the following definitions of various concepts.

\subsubsection{Task}
There are $m+1$ users, Users $1$, $2$, \ldots, $m$, $m+1$, and server(s).
When several server(s) exist, they are not allowed to communicate with each other.
%, Servers A and B. 
%Severs A and B are not allowed to communicate with each other.
Each user can communicate with both servers via classical channel.
Only server(s) are allowed to make quantum operations. 

User $j$ has the $j$-th component $(\bx_{j},\by_{j},\bz_{j})
\in \ZZ_4^n \times \ZZ_8^n \times \ZZ_2^{[n]_2}$ of the program 
$\bw$ for $j=1,2, \ldots, m$ while the server(s) do not know it.
The final user $m+1$ aims to get the initial $n_\circ$ bits of the computation outcome 
when the measurement based on the computation basis is performed
to the state $\bW(\bw)|0\rangle^{\otimes n} \in {\cal K} $.

When we apply this general method to a specific function $f$ with $l$ inputs 
$X_1,\ldots, X_l$,
we consider that there exist $l$ players who have respective inputs.
Then, the $m$ users are divided into $l$ distinct groups.
When the $k$-th group is composed of users with labels in the subset 
$S_k$,
the components $(\bx_{j},\by_{j},\bz_{j})_{j \in S_k}$ are decided by the choice of 
the value $X_k$.
That is, all users in the $k$-th group are controlled by 
$k$-th player.
When $l$ input variables $X_1,\ldots, X_l$ are independent of each other,
the information possessed by users in the $k$-th group
is independent of 
the information possessed by users in the $k'$-th group for $k\neq k'$.

\subsubsection{Example}
To see how the above general setting works in specific functions,
as an example, %for delegated multiparty quantum computation,
we consider the function 
$X_1+X_2+\cdots+X_l \in \ZZ_2$, where
each variable $X_j$ is an element of $\ZZ_2$.
The outcome of this function can be written as 
the final state of the following combination of universal gate set.
Consider a two-qubit system ${\cal H}_1\otimes {\cal H}_2$.
\begin{align}
&|X_1+X_2+\cdots+X_l \rangle_1 |1\rangle_2 \nonumber\\
\doteq
&(\sH_1 \sCZ_{(1,2)} \sT_1^4\sH_1\sT_1^4)^{X_l}
\cdots  (\sH_1 \sCZ_{(1,2)} \sT_1^4\sH_1\sT_1^4)^{X_2} \nonumber \\
&\cdot
(\sH_1 \sCZ_{(1,2)} \sT_1^4\sH_1\sT_1^4)^{X_1}
|0\rangle_1 |1\rangle_2 \nonumber\\
\doteq &
(\sH_1 \sCZ_{(1,2)} \sT_1^4\sH_1\sT_1^4)^{X_l}
\cdots  (\sH_1 \sCZ_{(1,2)} \sT_1^4\sH_1\sT_1^4)^{X_2} \nonumber \\
&\cdot
(\sH_1 \sCZ_{(1,2)} \sT_1^4\sH_1\sT_1^4)^{X_1}
\sH^2_2\sZ_2
|0\rangle_1 |0\rangle_2 \nonumber\\
= &
\sH_1^{X_l} \sCZ_{(1,2)}^{X_l} \sT_1^{4X_l} \sH_1^{X_l}\sT_1^{4X_l}
\cdots \sH_1^{X_2} \sCZ_{(1,2)}^{X_2} \sT_1^{4X_2}\sH_1^{X_2}\sT_1^{4X_2}
\nonumber \\
&\cdot
\sH_1^{X_1} \sCZ_{(1,2)}^{X_1} \sT_1^{4X_1}\sH_1^{X_1}\sT_1^{4X_1}
\sH^2_2 \sT_2^4
|0\rangle_1 |0\rangle_2 .
\end{align}
We set $m$ and $n_{\circ}$ to be $2l+1$ and $1$, respectively.
The first group $S_1$ is set to be $\{1,2,3\}$, and
the $k$-th group $S_k$ is set to be $\{2k,2k+1\}$ for 
$k=2,3, \ldots, l$.
Since $[2]_2$ is composed of one element $(1,2)$,
$\bz_j$ is composed of one element of $\ZZ_2$ for $j=1, \ldots, m=2l+1$.
Then, the program is chosen as
\begin{align}
(\bx_{1},\by_{1},\bz_{1})=& ( (0,2),(0,4),0)\\
(\bx_{2k},\by_{2k},\bz_{2k})=& ( (X_k,0),(4X_k,0),0)\\
(\bx_{2k+1},\by_{2k+1},\bz_{2k+1})=& ( (X_k,0),(4X_k,0),X_k)
\end{align}
for $k=1, \ldots, l$.
The measurement outcome with the computation basis 
on the first qubit in the final state is 
$X_1+X_2+\cdots+X_l \in \ZZ_2$.
Hence, the above choice of the program works for 
the delegated multiparty quantum computation of the function
$X_1+X_2+\cdots+X_l \in \ZZ_2$.

\begin{figure}[tb]{}
\begin{center} 
\includegraphics[width=\linewidth]{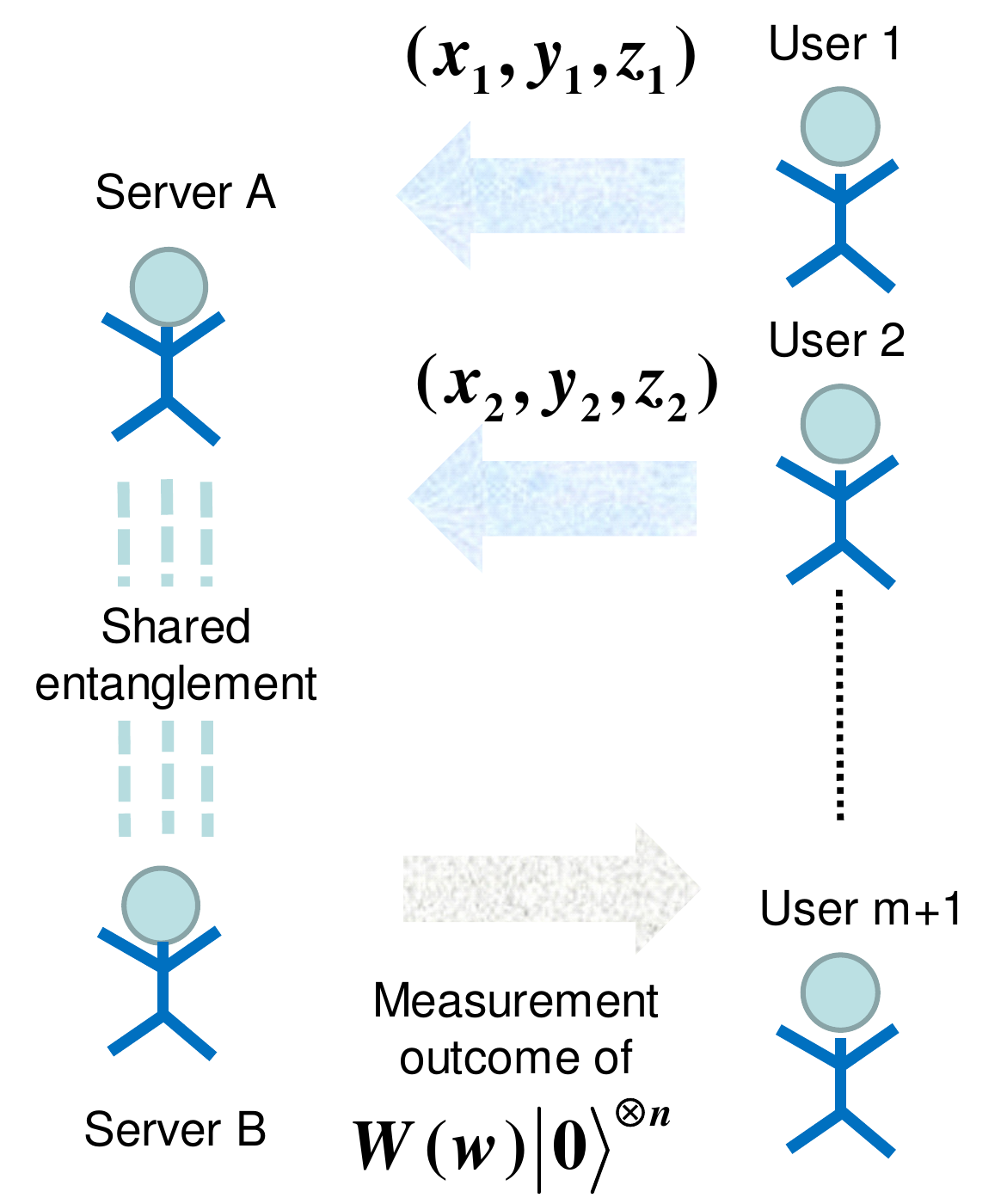}
\end{center}
\caption{{\bf Delegated multiparty quantum computation with two servers.}
This figure shows a protocol for delegated multiparty quantum computation
with two servers when the two servers share an entangled state.
}\Label{F2} 
\end{figure}

\subsubsection{Correctness and complexity}
Since a DMQC protocol $\Phi$ has bilateral communication.
The upload communication is the communication from the users to the server(s),
and  
the download communication is the communication from the server(s) to the users.
In the same way as OQC,
the communication complexity is composed of 
the upload complexity and the download complexity.

A DMQC protocol $\Phi$ is called correct when 
the following condition holds for any program
$\bw\in 
(\ZZ_4^n \times \ZZ_8^n \times \ZZ_2^{[n]_2} )^{m}$;
The outcome observed by the final user, User $m+1$,
is subject to 
the distribution of the initial $n_\circ$ bits of the measurement outcome
of the computation-basis measurement over 
the state $\bW(\bw)|0\rangle^{\otimes n} \in {\cal K} $.

\subsubsection{Secrecy condition}
We require the following secrecy conditions.
We choose an arbitrary subset $\Theta\subset [m]$ as a form 
$\cup_{k \in \bar{\Theta}} S_k$ with $\bar{\Theta} \subset [l] $.
(1)
When a set of users $\Theta \subset [m]$ is honest, 
server(s) obtain no information for 
$\{(\bx_{j},\by_{j},\bz_{j})\}_{j \in\Theta}$.
(2)
When Server(s) are honest additionally,
User $j'$ obtains no information for 
$\{(\bx_{j},\by_{j},\bz_{j})\}_{j \in\Theta\setminus \{j'\}}$ for $j'=1, \ldots, m$,
and the final user $m+1$ obtains no information for 
$\{(\bx_{j},\by_{j},\bz_{j})\}_{j \in\Theta}$ except for the above computation outcome, i.e.,
the final user $m+1$ can recover the final state on his/her own system
by using the above computation outcome and 
classical information generated by himself/herself.
(3) In addition, even when users in $[m+1]\setminus \Theta$ collude,
they obtain no information except for the above computation outcome.
When the above three conditions hold,
a DMQC  protocol $\Phi$ is called secure.

\begin{remark}
The paper \cite{KKMO} studied 
delegated multiparty quantum computation.
But, it allows the users to use single qubit operations and quantum communications.
Since our delegated multiparty quantum computation
does not allow the users to use single qubit operations nor quantum communications.

In addition, 
oblivious quantum computation cannot be considered as a special case of 
delegated multiparty quantum computation
in the sense of our definition nor 
in the sense of the definition by the paper \cite{KKMO} 
because of the following two reasons.
First,
the server(s) do not have input data in delegated multiparty quantum computation,
but they have program $\bw$ in oblivious quantum computation.
Second, oblivious quantum computation 
allows the user to input a quantum state and to obtain 
a quantum state as the output, but
delegated multiparty quantum computation does not allow them.
Therefore, 
oblivious quantum computation can be considered as a new concept.
\end{remark}

\subsection{Formulation of generalized delegated multiparty quantum computation}\Label{S43}
Delegated multiparty quantum computation is generalized to 
generalized delegated multiparty quantum computation (GDMQC)
by combining oblivious quantum computation as Fig. \ref{F3}
in the following way.

\subsubsection{Task}
There are $m+1$ users, Users $1$, $2$, \ldots, $m$, $m+1$, and server(s).
When several server(s) exist, they are not allowed to communicate with each other.
Each user can communicate with both servers via classical channel.
Only server(s) are allowed to make quantum operations. 
User $j$ has the $j$-th component $(\bx_{j}',\by_{j}',\bz_{j}')$ of a program 
$\bw'$ for $j=1,2, \ldots, m$ while the server(s) do not know it.
Server(s) have another program $\bw$ while the users do not know it.
The final user, User $m+1$, aims to get the initial $n_\circ$ bits of the computation outcome 
when the measurement based on the computation basis is performed
to the state $\bW(\bw\cdot\bw')|0\rangle^{\otimes n} \in {\cal K} $.

When we apply this general method to a specific function $f$ with $l$ inputs 
$X_1,\ldots, X_l,X_{l+1}$,
we consider that there exist $l$ players who have respective inputs.
Then, the $m$ users are divided into $l$ distinct groups.
When the $k$-th group is composed of users with labels in the subset 
$S_k$,
the components $(\bx_{j}',\by_{j}',\bz_{j}')_{j \in S_k}$ are decided by the choice of 
the value $X_k$.
That is, all users in the $k$-th group is controlled by $k$-th player.
Also, the final variable $X_{l+1}$ decides 
Server's program $\bw$.

\begin{figure}[tb]{}
\begin{center} 
\includegraphics[width=\linewidth]{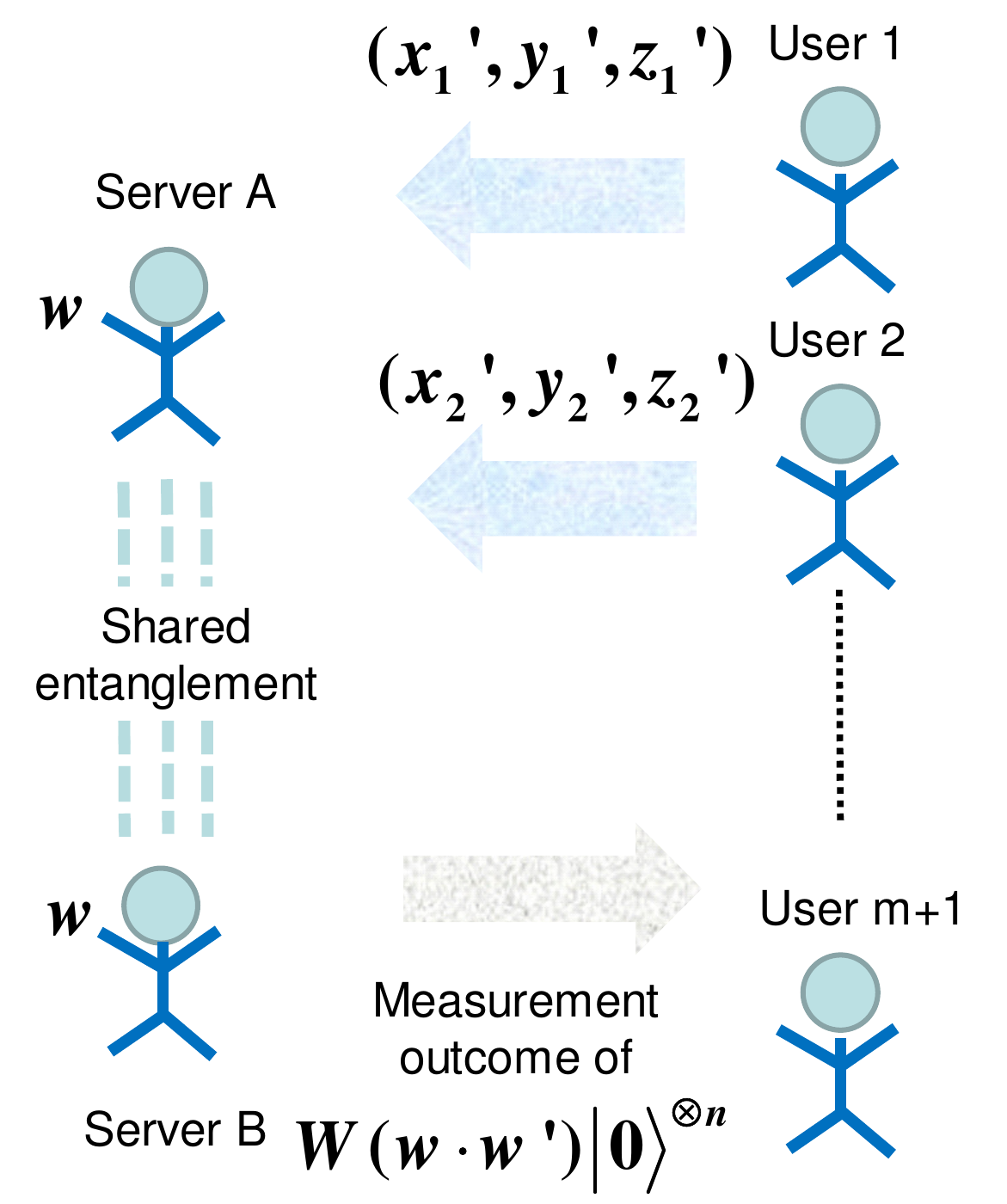}
\end{center}
\caption{{\bf Generalized delegated multiparty quantum computation with two servers.}
This figure shows a protocol for generalized delegated multiparty quantum computation
with two servers when the two servers share an entangled state.
}\Label{F3} 
\end{figure}

\subsubsection{Correctness and complexity}
For a GDMQC protocol $\Phi$,
its upload communication, its download communication,
and its communication complexity are defined in the same way 
as a DMQC protocol.

A DMQC protocol $\Phi$ is called correct when 
the following condition holds for any program
$\bw\in 
(\ZZ_4^n \times \ZZ_8^n \times \ZZ_2^{[n]_2} )^{m}$;
The outcome observed by the final user, User $m+1$,
is subject to 
the distribution of the initial $n_\circ$ bits of the measurement outcome
of the computation-basis measurement over 
the state $\bW(\bw)|0\rangle^{\otimes n} \in {\cal K} $.

\subsubsection{Two types of secrecy}
A GDMQC  protocol $\Phi$ has two types of secrecy.
One is the user-secrecy and the other is the server-secrecy.
We say that 
a GDMQC  protocol $\Phi$ satisfies the user-secrecy 
when the following conditions hold. 
We choose an arbitrary subset $\Theta\subset [m]$ as a form 
$\cup_{k \in \bar{\Theta}} S_k$ with $\bar{\Theta} \subset [l] $.
(1)
When  a set of users $\Theta \subset [m]$ are honest,
server(s) obtain no information for 
$\{(\bx_{j}',\by_{j}',\bz_{j}')\}_{j \in\Theta}$.
(2) When Server(s) are honest additionally,
User $j'$ obtains no information for 
$\{(\bx_{j}',\by_{j}',\bz_{j}')\}_{j \in\Theta\setminus \{j'\}}$ for $j'=1, \ldots, m$.
The final user $m+1$ obtains no information for 
$\{(\bx_{j}',\by_{j}',\bz_{j}')\}_{j \in\Theta}$ except for the above computation outcome, i.e,
the final user $m+1$ can recover the final state on his/her own system
by using the required computation outcome and 
classical information generated by himself/herself.
(3) In addition, even when users in $[m+1]\setminus \Theta$ collude,
they obtain no information except for the above computation outcome, i.e.,
they can recover the final state on their own system
by using the above computation outcome and 
classical information generated by themselves.

We say that an OQC  protocol $\Phi$ satisfies the server-secrecy 
when the following conditions hold. 
(1) When the server(s) are honest,
User $j$ with $j=1, \ldots, m$ obtains no information for $\bw$, 
and 
User $m+1$ obtains no information for $\bw$ expect to the required computation outcome, i.e.,
User $m+1$ can recover the final state on his/her own system
by using the required computation outcome and 
classical information generated by themselves.
(2) In addition, 
when the server(s) are honest,
even when all users collude,
the users obtain no information except for the above computation outcome, i.e., 
the users can recover the final state on their own system
by using the above computation outcome and 
classical information generated by themselves.

\subsubsection{Relation to OQC}\Label{S434}
A special case of GDMQC gives a variant of OQC as follows.
We divide $m$ rounds into two parts, the first $m_1$ rounds
and the remaining $m_2$ rounds, where $m=m_1+m_2$.
The first $m_1$ components of $\bw$ are set to be 
$\be_1:=(\underbrace{(1,1,1), \ldots,(1,1,1)}_{m_1})$,
and 
the remaining $m_2$ components of $\bw$ are denoted by $\bw_2$
Also,
the first $m_1$ components of $\bw'$ are denoted by $\bw_2'$, and
the remaining $m_2$ components of $\bw'$ are set to be 
$\be_2:=(\underbrace{(1,1,1), \ldots,(1,1,1)}_{m_2})$.
Also, we assume that there exists only one user. 

In this case, the user decides the state $\bW_1(\bw_1')|0\rangle^{\otimes n}$,
and the servers decide the unitary $\bW_2(\bw_2)$.
Finally, the user obtains the measurement outcome with the computation basis 
under the state $\bW_2(\bw_2)\bW_1(\bw_1')|0\rangle^{\otimes n}$.
When the user's input state is given as $\bW_1(\bw_1')|0\rangle^{\otimes n}$
and the user needs only the measurement outcome with the computation basis
in OQC,
this requirement can be done without quantum communication by using
a special case of GDMQC.

\section{Main protocol for TOQC}\Label{S7}
Since oblivious transfer with one server is impossible even with the quantum setting \cite{Mayers,Lo}, 
this section introduces a two-server  
OQC (TOQC) protocol that satisfies all the requirements. 
%Since Protocol \ref{PmainB} does not satisfy the user-secrecy with dishonest servers, but satisfies other requirements,
%we need to modify it to satisfy the user-secrecy with dishonest servers.
%For this aim, we attach a randomization for the input state to Protocol \ref{PmainB}.
\if0
We define
${\cal K}_{AB}:={\cal H}[\{|e_0\rangle, |e_1\rangle\}]^{\otimes n}$ and
${\cal K}_{AB,*}:={\cal H}[\{|e_0^*\rangle, |e_1^*\rangle\}]^{\otimes n}$,
and denote the projections to
${\cal K}_{AB}$ and ${\cal K}_{AB,*}$
by 
$P_{AB}$ and $P_{AB,*}$, respectively.
In the following except for Section \ref{S5P}, 
we set $|e_0\rangle:= |0\rangle$ and $|e_1\rangle:=e^{i\pi/4} |1\rangle$,
and simplify 
$\sS[|e_0\rangle, |e_1\rangle]$ and $\sR[|e_0\rangle, |e_1\rangle]$
to $\sS$ and $\sR$.

As shown in Section \ref{S6}, using the operator 
$\bar{\sH}:= e^{i \pi/4}|e_0^*\rangle \langle e_0^*|
+e^{-i \pi/4} |e_1^*\rangle \langle e_1^*|$,
we have 
\begin{align}
\begin{split}
\hat{V}  \sT &=(\sT \otimes I) \hat{V}  =(I \otimes \sT) \hat{V} ,\\
\hat{V}^{\otimes 2}  \sCZ 
&=(\sCZ \otimes I) \hat{V}^{\otimes 2}  =(I \otimes \sCZ) \hat{V}^{\otimes 2}  ,\end{split}
\Label{CNAO3}\\
\begin{split}
U[|e_0\rangle, |e_1\rangle] \hat{V}  \sH
&=(\bar{\sH}\otimes I)
U[|e_0\rangle, |e_1\rangle] \hat{V}  \\
&=(I \otimes \bar{\sH})
U[|e_0\rangle, |e_1\rangle] \hat{V}  .
\end{split}
\Label{CNAO}
\end{align}
Hence, the applications of $\sT,\sCZ,\sH$ can be replaced by the applications of 
$\sT,\sCZ,\bar{\sH}$ after $\hat{\sV}$ or $U[|e_0\rangle, |e_1\rangle] \hat{V}$.
\fi
Then, we introduce the following protocol.
 \begin{prot}[Main protocol for TOQC] \Label{Pmain}
%The following TOQC protocol with server-secrecy and user-secrecy
Severs A and B are not allowed to communicate with each other.
The user can communicate with both servers.
%Our protocol is given as follows.
\begin{description}[leftmargin=1.5em]
\item[0)] \textbf{Entanglement Sharing}: 
	Servers A and B prepare $2mn$ copies of the Bell state $|\Phi\rangle$
on ${\cal H}_{A,j,s}\otimes {\cal H}_{B,j,s}$ for 
$s=1, \ldots, n$ and $j=1, \ldots, 2m$.
\item[1)] \textbf{Query 1}: %The same as Protocol \ref{PR1}.
The user generates a quantum state $|\psi\rangle \in {\cal H}^{\otimes n}
=\otimes_{s=1}^n {\cal H}_s $, 
and random bits $A_{0,s},B_{0,s} \in\ZZ_2$ 
according to the uniform distribution
for $s=1, \ldots, n$, where
${\cal H}_s$ is the $s$-th system.
The user applies $\sZ_s^{B_{0,s}} \sX_s^{A_{0,s}}$ on ${\cal H}_s$, 
and obtains
the state $\otimes_{s=1}^n \sZ_s^{B_{0,s}} \sX_s^{A_{0,s}} 
|\psi\rangle \in {\cal H}^{\otimes n} $.
Then, the user sends the system ${\cal H}^{\otimes n} $ to Server A.
Also,		the user generates
				$\bQ_{2,1,u}=(Q_{2,1,u,s})_{s \in [n]}  \in  \ZZ_8^{n}$ 
				and
				$\bQ_{3,1,(u,v)}=(Q_{3,1,(u,v),(s,t)})_{(s,t) \in [n]_2}
				\in  \ZZ_2^{[n]_2}$ 
				uniformly at random for $(u,v) \in \ZZ_2^2$, and sends them to Server A.

\item[2)] \textbf{Unitary operation 1}: 
Server A applies unitaries
$\sX_s^u \sT_s(Q_{2,1,u,s} \by_{1,s}) \sX_s^u$ to ${\cal H}_s$
and $\sX_s^u \sX_t^v
 \sCZ_{(s,t)}(Q_{3,1,(u,v),(s,t)} \bz_{1,(s,t)}) \sX_s^u \sX_t^v$ 
to ${\cal H}_{s}\otimes {\cal H}_{t}$.
Server A applies the Bell measurement 
$\{|\Phi_{a,b}\rangle\}_{a,b \in \ZZ_2}$
on ${\cal H}_s\otimes {\cal H}_{A,1,s}$ for $s=1, \ldots, n$.
Then, Server A obtains the outcome $(A_{1,s},B_{1,s})$ and sends them to the user
for $s=1, \ldots, n$.
Server A applies unitary
$\sZ_s^{B_{1,s}} \sX_s^{A_{1,s}}$ to ${\cal H}_{A,2,s}$.

\item[3)] \textbf{Query 2}: 
The user generates
				$\bQ_{2,1,u}'=(Q_{2,1,u,s}')_{s \in [n]}  \in  \ZZ_8^{n}$ 
				and
				$\bQ_{3,1,(u,v)}'=(Q_{3,1,(u,v),(s,t)}')_{(s,t) \in [n]_2}
				\in  \ZZ_2^{[n]_2}$ as
\begin{align}
Q_{2,1,u,s}' &:=-Q_{2,1,u- A_{1,s},s}
+\delta_{u,A_{0,s}+A_{1,s}} ,\Label{BG1}
\end{align}
and
\begin{align}
&Q_{3,1,(u,v),(s,t)}' \nonumber \\
&:=-Q_{3,1,(u- A_{1,s},v- A_{1,t}),(s,t)}
+\delta_{u,A_{0,s}+A_{1,s}} \delta_{v,A_{0,t}+A_{1,t}} .
\Label{BG2}
\end{align}
\if0
\begin{align}
Q_{2,1,A_{1,s}+A_{2,s},s}' &:=-Q_{2,1,A_{1,s},s}+1 \\
Q_{2,1,A_{1,s}+A_{2,s}+1,s}'&:=-Q_{2,1,A_{1,s}+1,s} \\
Q_{3,1,(A_{1,s}+A_{2,s},A_{1,t}+A_{2,t}),(s,t)}' 
&:=-Q_{3,1,(A_{1,s},A_{1,t}),(s,t)}+1 \\
Q_{3,1,(A_{1,s}+A_{2,s}+1,A_{1,t}+A_{2,t}),(s,t)}' 
&:=-Q_{3,1,(A_{1,s}+1,A_{1,t}),(s,t)}+1 \\
Q_{3,1,(A_{1,s}+A_{2,s},A_{1,t}+A_{2,t}+1),(s,t)}' 
&:=-Q_{3,1,(A_{1,s},A_{1,t}+1),(s,t)}+1 \\
Q_{3,1,(A_{1,s}+A_{2,s}+1,A_{1,t}+A_{2,t}+1),(s,t)}' 
&:=-Q_{3,1,(A_{1,s}+1,A_{1,t}+1),(s,t)}+1 .
\end{align}
\fi
The user generates
				$\bQ_{1,1,u}'=(Q_{1,1,u,s}')_{s \in [n]}  \in  \ZZ_4^{n}$ 
according to the uniform distribution. 
The user sends them to Server B.

\item[4)] \textbf{Unitary operation 2}: 
(i) Server B applies unitaries
$\sX_s^u \sT_s(Q_{2,1,u,s}' \by_{1,s}) \sX_s^u$ to ${\cal H}_{B,1,s}$
and $\sX_s^u \sX_t^v
 \sCZ_{(s,t)}(Q_{3,1,(u,v),(s,t)}' \bz_{1,(s,t)}) \sX_s^u \sX_t^v$ 
to ${\cal H}_{B,1,s}\otimes {\cal H}_{B,1,t}$ for $(u,v) \in \ZZ_2^2$.
(ii) Then,
Server B applies unitaries
$\sX_s^u \sH_s(Q_{1,1,u,s}' \by_{1,s}) \sX_s^u$ to ${\cal H}_{B,1,s}$
for $u \in \ZZ_2$.
(iii) Server B applies the Bell measurement 
$\{|\Phi_{a,b}\rangle\}_{a,b \in \ZZ_2}$
on ${\cal H}_{B,1,s}\otimes {\cal H}_{B,2,s}$ for $s=1, \ldots, n$.
Then, Server B obtains the outcome $(A_{2,s},B_{2,s})$ and sends them to the user
for $s=1, \ldots, n$.
(iv) Server B applies unitary
$\sZ_s^{B_{2,s}} \sX_s^{A_{2,s}}$ to ${\cal H}_{B,3,s}$.
\end{description}
\noindent {\bf We perform the following steps for $j=2,\ldots,m $.}
\begin{description}
\item[4$j$-3)] \textbf{Query 2$j$-1}: 
The user generates
				$\bQ_{1,j-1,u}=(Q_{1,j-1,u,s})_{s \in [n]}  \in  \ZZ_4^{n}$ as
\begin{align}
&Q_{1,j-1,u,s} \nonumber \\
:=&-Q_{1,j-1,u- (A_{2j-2,s}+B_{2j-2,s}+A_{2j-3,s}+B_{2j-3,s}),s}' \nonumber \\
&+\delta_{u,
A_{0,s}+B_{0,s}+A_{2j-2,s}+B_{2j-2,s}} \Label{ZMT}.
\end{align}
\if0
\begin{align}
Q_{1,j-1,\sum_{k=1}^{2j-1}A_{k,s}+B_{k,s},s }&
:=-Q_{1,j-1,\sum_{k=1}^{2j-2}A_{k,s}+B_{k,s},s}'+1 \\
Q_{1,j-1,(\sum_{k=1}^{2j-1}A_{k,s}+B_{k,s})+1,s}&
:=-Q_{1,j-1,(\sum_{k=1}^{2j-2}A_{k,s}+B_{k,s})+1,s}' .
\end{align}
\fi
Also, the user generates
				$\bQ_{2,j,u}=(Q_{2,j,u,s})_{s \in [n]}  \in  \ZZ_8^{n}$ 
				and
				$\bQ_{3,j,(u,v)}=(Q_{3,j,(u,v),(s,t)})_{(s,t) \in [n]_2}
				\in  \ZZ_2^{[n]_2}$ 
				uniformly at random for $(u,v) \in \ZZ_2^2$.
		Then, the user sends them to Server A. 

\item[4$j$-2)] \textbf{Unitary operation  2$j$-1}: 
(i) Server A applies unitaries
$\sX_s^u \sH_s(Q_{1,j-1,u,s} \by_{j-1,s}) \sX_s^u$ to ${\cal H}_{A,2j-2,s}$
for $u \in \ZZ_2$.
(ii) Then, 
Server A applies unitaries
$\sX_s^u \sT_s(Q_{2,j,u,s} \by_{j,s}) \sX_s^u$ to ${\cal H}_{A,2j-2,s}$
and $\sX_s^u \sX_t^v
 \sCZ_{(s,t)}(Q_{3,j,(u,v),(s,t)} \bz_{j,(s,t)}) \sX_s^u \sX_t^v$ 
to ${\cal H}_{A,2j-2,s}\otimes {\cal H}_{A,2j-2,t}$.
(iii) Server A applies the Bell measurement 
$\{|\Phi_{a,b}\rangle\}_{a,b \in \ZZ_2}$
on ${\cal H}_{A,2j-2,s}\otimes {\cal H}_{A,2j-1,s}$ for $s=1, \ldots, n$.
(iv) Then, Server A obtains the outcome $(A_{2j-1,s},B_{2j-1,s})$ and sends them to the user
for $s=1, \ldots, n$.
Server A applies unitary
$\sZ_s^{B_{2j-1,s}} \sX_s^{A_{2j-1,s}}$ to ${\cal H}_{A,2j,s}$.

\item[4$j$-1)] \textbf{Query 2$j$}: 
The user generates
				$\bQ_{2,j,u}'=(Q_{2,j,u,s}')_{s \in [n]}  \in  \ZZ_8^{n}$ 
				and
				$\bQ_{3,j,(u,v)}'=(Q_{3,j,(u,v),(s,t)}')_{(s,t) \in [n]_2}
				\in  \ZZ_2^{[n]_2}$ as
\begin{align}
Q_{2,j,u,s}' &:=-Q_{2,j,u- A_{2j-1,s}-A_{2j-2,s},s}
+\delta_{u,A_{0,s}+A_{2j-1,s}} \Label{BG3}
\end{align}
and
\begin{align}
&Q_{3,j,(u,v),(s,t)}' \nonumber \\
:=&-Q_{3,j,(u- A_{2j-1,s}-A_{2j-2,s},v- A_{2j-1,t}-A_{2j-2,t}),(s,t)}\nonumber \\
&+\delta_{u,A_{0,s}+A_{2j-1,s}} 
\delta_{v,A_{0,t}+A_{2j-1,t}} .\Label{BG4}
\end{align}
\if0
\begin{align}
Q_{2,j,\sum_{k=1}^{2j}A_{k,s},s}' &:=-Q_{2,j,\sum_{k=1}^{2j-1}A_{k,s},s}+1 \\
Q_{2,j,(\sum_{k=1}^{2j}A_{k,s})+1,s}'&:=-Q_{2,j,(\sum_{k=1}^{2j-1}A_{k,s})+1,s} \\
Q_{3,j,(A_{1,s}+A_{2,s},A_{1,t}+A_{2,t}),(s,t)}' 
&:=-Q_{3,j,(A_{1,s},A_{1,t}),(s,t)}+1 \\
Q_{3,1,(A_{1,s}+A_{2,s}+1,A_{1,t}+A_{2,t}),(s,t)}' 
&:=-Q_{3,j,(A_{1,s}+1,A_{1,t}),(s,t)}+1 \\
Q_{3,j,(A_{1,s}+A_{2,s},A_{1,t}+A_{2,t}+1),(s,t)}' 
&:=-Q_{3,j,(A_{1,s},A_{1,t}+1),(s,t)}+1 \\
Q_{3,j,(A_{1,s}+A_{2,s}+1,A_{1,t}+A_{2,t}+1),(s,t)}' 
&:=-Q_{3,j,(A_{1,s}+1,A_{1,t}+1),(s,t)}+1 .
\end{align}
\fi
The user generates
				$\bQ_{1,j,u}'=(Q_{1,j,u,s}')_{s \in [n]}  \in  \ZZ_4^{n}$ 
according to the uniform distribution. 
The user sends them to Server B.

\item[4$j$)] \textbf{Unitary operation 2$j$}: 
(i) Server B applies unitaries
$\sX_s^u \sT_s(Q_{2,j,u,s}' \by_{j1,s}) \sX_s^u$ to ${\cal H}_{B,2j-1,s}$
and $\sX_s^u \sX_t^v
 \sCZ_{(s,t)}(Q_{3,j,(u,v),(s,t)}' \bz_{j,(s,t)}) \sX_s^u \sX_t^v$ 
to ${\cal H}_{B,2j-1,s}\otimes {\cal H}_{B,2j-1,t}$ for $(u,v) \in \ZZ_2^2$.
(ii) Then,
Server B applies unitaries
$\sX_s^u \sH_s(Q_{1,j,u,s}' \by_{j,s}) \sX_s^u$ to ${\cal H}_{B,2j-1,s}$
for $u \in \ZZ_2$.
(iii) Server B applies the Bell measurement 
$\{|\Phi_{a,b}\rangle\}_{a,b \in \ZZ_2}$
on ${\cal H}_{B,2j-1,s}\otimes {\cal H}_{B,2j,s}$ for $s=1, \ldots, n$.
Then, Server B obtains the outcome $(A_{2j,s},B_{2j,s})$ and sends them to the user
for $s=1, \ldots, n$.
(iv) Server B applies unitary
$\sZ_s^{B_{2j,s}} \sX_s^{A_{2j,s}}$ to ${\cal H}_{B,2j+1,s}$.
\end{description}

\noindent {\bf After Steps with $j=m$, we perform the following remaining steps.}

\begin{description}
\item[4$m$+1)] \textbf{Query 2$m$+1}: 
The user generates
				$\bQ_{1,m,u}'=(Q_{1,m,u,s}')_{s \in [n]}  \in  \ZZ_4^{n}$ as \eqref{ZMT}
				with $j=m+1$.
		Then, the user sends them to Server A. 

\item[4$m$+2)] \textbf{Unitary operation  2$m$+1}: 
(i) Server A applies unitaries
$\sX_s^u \sH_s(Q_{1,m,u,s} \by_{m,s}) \sX_s^u$ to ${\cal H}_{A,2m,s}$
for $u \in \ZZ_2$.
(ii) Then, Server A sends the system ${\cal H}_{A,2m,s}$ for $s=1, \ldots, n_{\circ}$ to the user.

\item[4$m$+3)] \textbf{Construction}:
The user applies 
$\sX_s^{A_{0,s}+A_{2m,s}}
\sZ_s^{B_{0,s}+B_{2m,s}}$ to ${\cal H}_{A,2m,s}$ for $s=1, \ldots, n_{\circ}$. 
\end{description}
\end{prot}

In the above protocol, 
the first basis conversion is done as Step 2) before the first unitary operation, Step 4).
Since the output state of Step 2) belongs to two-dimensional subspace,
the problem (b) is resolved.
This is the reason why the first unitary operation is not done before the first 
basis conversion.

\begin{theo}\Label{Th2}
Protocol \ref{Pmain} is a correct TOQC protocol that satisfies
the user-secrecy even with dishonest servers and 
the server-secrecy even with a dishonest user.
Its upload complexity is $(4n^2+16n)m$ bits and $n$ qubits.
Its download complexity is $4nm$ bits and $n_\circ$ qubits
\end{theo}

Although Step 1) of Protocol \ref{Pmain} contains quantum communications,
when the state $|\psi\rangle$ is limited to basis states for the basis 
$\{ |\bx\rangle\}_{\bx \in \ZZ_2^n}$,
the quantum communication in Step 1) can be replaced to classical communication
because 
the masked state is also a basis state.
In addition, when the user applies the measurement based on the computation basis
in Step $4m+3$),
Steps $4m+2$) and $4m+3$) can be replaced by 
the following Steps;
\begin{description}
\item[4$m$+2')] \textbf{Unitary operation  2$m$+1}: 
(i) Server A applies unitaries
$\sX_s^u \sH_s(Q_{1,m,u,s} \by_{m,s}) \sX_s^u$ to ${\cal H}_{A,2m,s}$
for $u \in \ZZ_2$.
(ii) Then, Server A 
applies the measurement $\{ |0\rangle, |1\rangle \}$
to the system ${\cal H}_{A,2m,s}$ and obtains the outcome
$X_s $
for $s=1, \ldots, n_{\circ}$.
(iii) Server A sends
the outcome $X_s $
for $s=1, \ldots, n_{\circ}$ to User $m+1$.

\item[4$m$+3')] \textbf{Construction}:
The user calculates 
$A_{2m,s}+X_{s}$ for $s=1, \ldots, n_{\circ}$. 
\end{description}
In this way, when the above two conditions are satisfied, 
this protocol does not need quantum communication.

\begin{proof}
\noindent{\bf Complexity}: 
First, we discuss its upload complexity.
In Step 1), the size of the transmitted quantum system is $n$ qubits.
$Q_{2,1,u,s}$ and $Q_{3,1,(u,v),(s,t)}$ have 3bits and 1 bit, respectively, for $(s,t) \in [n]_2$ and $(u,v) \in \ZZ_2^2 $. 
Hence, $\bQ_{2,1,u}$ and $\bQ_{3,1,(u,v)}$ have $3n$ bits and $n(n-1)/2$ bits, respectively, for $(u,v) \in \ZZ_2^2 $. 
Thus, in Step 1), the user sends $2 \cdot 3n+4 \cdot n(n-1)/2= 
6n+2n(n-1)=2n^2+4n$ bits in addition to $n$ qubits
to Server A.
In Step 3),
since $Q_{1,1,u,s}'$ has 2bits for $s \in [n]$ and $u \in \ZZ_2$,
$\bQ_{1,1,u}'$ has $2n$ bits for $u \in \ZZ_2$. 
Thus, in Step 3), 
since the sender sends
$\bQ_{2,1,u}'$, $\bQ_{3,1,(u,v)}'$, and $\bQ_{1,1,u}'$
to Server B, 
Step 3) has upload complexity $2n^2+8n$
In the same way, 
Steps 4$j$-3) and 4$j$-1) have upload complexity $2n^2+8n$.
Step 4$m$+1) has upload complexity $4n$ to send $\bQ_{1,m,u}$ for $u \in \ZZ_2$. 
Hence, the total upload complexity $(4n^2+16n)m$ bits and $n$ qubits.

Next, we discuss its download complexity.
Steps 2), 4), 4$j$-2), and 4$j$) have download complexity with $2n$ bits.
Step 4$m$+2) has download complexity with $n_{\circ}$ qubits.
Hence, the total download complexity $4n m$ bits and $n_{\circ}$ qubits.

\noindent{\bf Correctness}: 
\begin{widetext}
The state at the end of (i) of Step 3) on $\otimes_{s=1}^n {\cal H}_{B,1,s}$
is the following.
\begin{align}
&|\psi_1\rangle\nonumber \\
&:=\Big(\bigotimes_{(s,t) \in [n]_2}
\prod_{(u,v) \in \ZZ_2^2}
\sX_s^u \sX_t^v
 \sCZ_{(s,t)}(Q_{3,1,(u,v),(s,t)}' \bz_{1,(s,t)}) \sX_s^u \sX_t^v
\Big)
\Big(\bigotimes_{s=1}^n
\prod_{u \in \ZZ_2}
\sX_s^u \sT_s(Q_{2,1,u,s}' \by_{1,s}) \sX_s^u\Big)
\Big(\bigotimes_{s=1}^n \sZ_s^{B_{1,s}} \sX_s^{A_{1,s}} \Big)\nonumber \\
&\quad\cdot\Big(\bigotimes_{(s,t) \in [n]_2}
\prod_{(u,v) \in \ZZ_2^2}
\sX_s^u \sX_t^v
 \sCZ_{(s,t)}(Q_{3,1,(u,v),(s,t)} \bz_{1,(s,t)}) \sX_s^u \sX_t^v
\Big)
\Big(\bigotimes_{s=1}^n
\prod_{u \in \ZZ_2}
\sX_s^u \sT_s(Q_{2,1,u,s} \by_{1,s}) \sX_s^u\Big)
\Big(\bigotimes_{s=1}^n \sZ_s^{B_{0,s}} \sX_s^{A_{0,s}} \Big)
|\psi\rangle\nonumber \\
&\doteq
\Big(\bigotimes_{s=1}^n \sZ_s^{B_{1,s}} \sX_s^{A_{1,s}} \Big)
\Big(\bigotimes_{s=1}^n \sZ_s^{B_{0,s}} \sX_s^{A_{0,s}} \Big)
\Big(\bigotimes_{s \in [n]}\sT_s(\by_{1,s})\Big)
\Big(\bigotimes_{(s,t) \in [n]_2} \sCZ_{(s,t)}( \bz_{1,(s,t)})\Big)
|\psi\rangle ,\Label{MVA}
\end{align}
where the above equation follows from 
\eqref{BG1} and \eqref{BG2}. 

%The state at the end of (i) of Step 5) is the following.
The state at the end of (i) of Step 5) on $\otimes_{s=1}^n {\cal H}_{A,2,s}$
is the following.
\begin{align} 
&|\psi_2\rangle\nonumber \\
&:=\Big(\bigotimes_{s=1}^n
\prod_{u \in \ZZ_2}
\sX_s^u \sH_s(Q_{1,1,u,s} \bx_{1,s}) \sX_s^u\Big)
\Big(\bigotimes_{s=1}^n \sZ_s^{B_{1,s}} \sX_s^{A_{1,s}} \Big)
\Big(\bigotimes_{s=1}^n \sZ_s^{B_{2,s}} \sX_s^{A_{2,s}} \Big)
\Big(\bigotimes_{s=1}^n
\prod_{u \in \ZZ_2}
\sX_s^u \sH_s(Q_{1,1,u,s}' \bx_{1,s}) \sX_s^u\Big)
|\psi_1\rangle\nonumber \\
&\doteq
\Big(\bigotimes_{s=1}^n
\prod_{u \in \ZZ_2}
\sX_s^u \sH_s(Q_{1,1,u,s} \bx_{1,s}) \sX_s^u\Big)
\Big(\bigotimes_{s=1}^n \sY_s^{B_{1,s}} \sX_s^{A_{1,s}+B_{1,s}} \Big)
\Big(\bigotimes_{s=1}^n \sY_s^{B_{2,s}} \sX_s^{A_{2,s}+B_{2,s}} \Big)
\Big(\bigotimes_{s=1}^n
\prod_{u \in \ZZ_2}
\sX_s^u \sH_s(Q_{1,1,u,s}' \bx_{1,s}) \sX_s^u\Big)
|\psi_1\rangle\nonumber \\
&\stackrel{(a)}{\doteq}
\Big(\bigotimes_{s=1}^n \sY_s^{B_{2,s}} \sX_s^{A_{2,s}+B_{2,s}} \Big)
\Big(\bigotimes_{s=1}^n \sY_s^{B_{0,s}} \sX_s^{A_{0,s}+B_{0,s}} \Big)
\Big(\bigotimes_{s \in [n]}\sH_s(\bx_{1,s})\Big)
\Big(\bigotimes_{s=1}^n \sY_s^{B_{0,s}} \sX_s^{A_{0,s}+B_{0,s}} \Big)
\Big(\bigotimes_{s=1}^n \sY_s^{B_{1,s}} \sX_s^{A_{1,s}+B_{1,s}} \Big)
|\psi_1\rangle\nonumber \\
&\doteq
\Big(\bigotimes_{s=1}^n \sZ_s^{B_{2,s}} \sX_s^{A_{2,s}} \Big)
\Big(\bigotimes_{s=1}^n \sZ_s^{B_{0,s}} \sX_s^{A_{0,s}} \Big)
\Big(\bigotimes_{s \in [n]}\sH_s(\bx_{1,s})\Big)
\Big(\bigotimes_{s=1}^n \sZ_s^{B_{0,s}} \sX_s^{A_{0,s}} \Big)
\Big(\bigotimes_{s=1}^n \sZ_s^{B_{1,s}} \sX_s^{A_{1,s}} \Big)
|\psi_1\rangle\nonumber \\
&\stackrel{(b)}{\doteq}
\Big(\bigotimes_{s=1}^n \sZ_s^{B_{2,s}} \sX_s^{A_{2,s}} \Big)
\Big(\bigotimes_{s=1}^n \sZ_s^{B_{0,s}} \sX_s^{A_{0,s}} \Big)
\Big(\bigotimes_{s \in [n]}\sH_s(\bx_{1,s})\Big)
\Big(\bigotimes_{s \in [n]}\sT_s(\by_{1,s})\Big)
\Big(\bigotimes_{(s,t) \in [n]_2} \sCZ_{(s,t)}( \bz_{1,(s,t)})\Big) 
|\psi\rangle ,\Label{MVA2}
\end{align}
where $(a)$ follows from \eqref{ZMT} and the relation $[\sY_s,\sH_s]=0$, 
and $(b)$ follows from \eqref{MVA}.

The state at the end of (i) of Step 7) on $\otimes_{s=1}^n {\cal H}_{B,3,s}$
is the following.
\begin{align}
&|\psi_3\rangle\nonumber \\
&:=\Big(\bigotimes_{(s,t) \in [n]_2}
\prod_{(u,v) \in \ZZ_2^2}
\sX_s^u \sX_t^v
 \sCZ_{(s,t)}(Q_{3,2,(u,v),(s,t)}' \bz_{2,(s,t)}) \sX_s^u \sX_t^v
\Big)
\Big(\bigotimes_{s=1}^n
\prod_{u \in \ZZ_2}
\sX_s^u \sT_s(Q_{2,2,u,s}' \by_{2,s}) \sX_s^u\Big)
\Big(\bigotimes_{s=1}^n \sZ_s^{B_{2,s}} \sX_s^{A_{2,s}} \Big)\nonumber \\
&\quad\cdot
\Big(\bigotimes_{s=1}^n \sZ_s^{B_{3,s}} \sX_s^{A_{3,s}} \Big)
\Big(\bigotimes_{(s,t) \in [n]_2}
\prod_{(u,v) \in \ZZ_2^2}
\sX_s^u \sX_t^v
 \sCZ_{(s,t)}(Q_{3,2,(u,v),(s,t)} \bz_{2,(s,t)}) \sX_s^u \sX_t^v
\Big)
\Big(\bigotimes_{s=1}^n
\prod_{u \in \ZZ_2}
\sX_s^u \sT_s(Q_{2,2,u,s} \by_{2,s}) \sX_s^u\Big)
|\psi_2\rangle\nonumber \\
&\stackrel{(a)}{\doteq}
\Big(\bigotimes_{s=1}^n \sZ_s^{B_{3,s}} \sX_s^{A_{3,s}} \Big)
\Big(\bigotimes_{s=1}^n \sZ_s^{B_{0,s}} \sX_s^{A_{0,s}} \Big)
\Big(\bigotimes_{s \in [n]}\sT_s(\by_{2,s})\Big)
\Big(\bigotimes_{(s,t) \in [n]_2} \sCZ_{(s,t)}( \bz_{2,(s,t)})\Big)
\Big(\bigotimes_{s=1}^n \sZ_s^{B_{2,s}} \sX_s^{A_{2,s}} \Big)
\Big(\bigotimes_{s=1}^n \sZ_s^{B_{0,s}} \sX_s^{A_{0,s}} \Big)
|\psi_2\rangle\nonumber  \\
&\stackrel{(b)}{\doteq}
\Big(\bigotimes_{s=1}^n \sZ_s^{B_{3,s}} \sX_s^{A_{3,s}} \Big)
\Big(\bigotimes_{s=1}^n \sZ_s^{B_{0,s}} \sX_s^{A_{0,s}} \Big)
\Big(\bigotimes_{s \in [n]}\sT_s(\by_{2,s})\Big)
\Big(\bigotimes_{(s,t) \in [n]_2} \sCZ_{(s,t)}( \bz_{2,(s,t)})\Big)\nonumber  \\
&\quad\cdot \Big(\bigotimes_{s \in [n]}\sH_s(\bx_{1,s})\Big)
\Big(\bigotimes_{s \in [n]}\sT_s(\by_{1,s})\Big)
\Big(\bigotimes_{(s,t) \in [n]_2} \sCZ_{(s,t)}( \bz_{1,(s,t)})\Big) 
|\psi\rangle ,
\end{align}
where $(a)$ follows from \eqref{BG3} and \eqref{BG4}, and $(b)$ follows from \eqref{MVA2}.

Repeating the above discussion, we find that 
the state at the end of (i) of Step 4$m$+2) on $\otimes_{s=1}^n {\cal H}_{A,2m,s}$
is the following.
\begin{align}
|\psi_{2m}\rangle
:=
\Big(\bigotimes_{s=1}^n \sZ_s^{B_{2m,s}} \sX_s^{A_{2m,s}} \Big)
\Big(\bigotimes_{s=1}^n \sZ_s^{B_{0,s}} \sX_s^{A_{0,s}} \Big)
\Bigg(\prod_{j=1}^m \Big(\bigotimes_{s \in [n]}\sH_s(\bx_{j,s})\Big)
\Big(\bigotimes_{s \in [n]}\sT_s(\by_{j,s})\Big)
\Big(\bigotimes_{(s,t) \in [n]_2} \sCZ_{(s,t)}( \bz_{j,(s,t)})\Big) 
\Bigg) |\psi\rangle .
\end{align}
Then, Step 4$m$+3) constructs the desired state from $|\psi_{2m}\rangle$.
\end{widetext}

\noindent{\bf Server-secrecy}: 
Next, we show the server-secrecy even with a dishonest user.
Assume that the servers are honest.
The user obtains the variables $(A_{k,s},B_{k,s})$ for $k=1, \ldots, 2m$ and $s=1,\ldots,n$,
and the quantum system $\otimes_{s=1}^{n_{\circ}}{\cal H}_{A,2m,s}$.
These variables are independent of the program $\bw$.
Hence, 
only the state on the quantum system $\otimes_{s=1}^{n_{\circ}}{\cal H}_{A,2m,s}$
is related to  the program $\bw$.
Since its dimension is the same as the desired output information, 
the user does not obtain any information for 
the program $\bw$ more than the desired output information.
In other words,
the user can generate the final state on his/her own whole system
by using $\Phi_{ideal}(\bw,|\psi\rangle)$,
the classical information describing the initial state $|\psi\rangle$,
and classical information generated by himself/herself.

\noindent{\bf User-secrecy}: 
Next, we show the user-secrecy even with dishonest servers.
Assume that the user is honest.
Server A receives the system $\otimes_{s=1}^n {\cal H}_{s}$ 
and the variables
$ \bQ_{1,j,u}$, $ \bQ_{2,j,u}$,
$\bQ_{3,j,(u,v)}$ for $j=1, \ldots,m$ and $(u,v)\in\ZZ_2^2$.
Since the state on the system $\otimes_{s=1}^n {\cal H}_{s}$ is the completely mixed state
and these variables are subject to the uniform distribution
independently.
Since they are independent of $|\psi\rangle$,
Server A obtains no information for  $|\psi\rangle$.
Also,
Server B receives 
the variables
$ \bQ_{1,j,u}'$, $ \bQ_{2,j,u}'$,
$\bQ_{3,j,(u,v)}'$ for $j=1, \ldots,m$ and $(u,v)\in\ZZ_2^2$.
These variables are subject to the uniform distribution
independently.
Since they are independent of $|\psi\rangle$,
Server B obtains no formation for  $|\psi\rangle$.
Therefore, 
the user-secrecy holds even with dishonest servers.

\end{proof}

\section{Protocol for TGDMQC}\Label{S8-2}
%Generalized Protocol without quantum communication}
Since one-server delegated quantum computation is impossible \cite{DK,MK,MNTT},
using Protocol \ref{Pmain},
we introduce our two-server GDMQC (TGDMQC) protocol  
by considering the case when the input is fixed to $|0\rangle^{\otimes n}$
and the read-out measurement is fixed to the measurement based on the computation basis.
In this case, the task can be done even when the user makes only classical communication.

\begin{prot}[TGDMQC protocol] \Label{Pro-C}
%The following TOQC protocol with server-secrecy and user-secrecy
There are $m+1$ users, Users $1$, $2$, \ldots, $m$, $m+1$, and two servers, Servers A and B. 
Severs A and B are not allowed to communicate with each other.
Each user can communicate with both servers with a classical channel.
Only Servers A and B are allowed to make quantum operations. 
%Our protocol is given as follows.
\begin{description}[leftmargin=1.5em]
\item[0)] \textbf{Entanglement Sharing}: 
	Servers A and B prepare $2mn$ copies of the Bell state $|\Phi\rangle$
on ${\cal H}_{A,j,s}\otimes {\cal H}_{B,j,s}$ for 
$s=1, \ldots, n$ and $j=1, \ldots, 2m$.
\item[1)] \textbf{Query 1}: %The same as Protocol \ref{PR1}.
User 1 generates
				$\bQ_{2,1,u}=(Q_{2,1,u,s})_{s \in [n]}  \in  \ZZ_8^{n}$ 
				and
				$\bQ_{3,1,(u,v)}=(Q_{3,1,(u,v),(s,t)})_{(s,t) \in [n]_2}
				\in  \ZZ_2^{[n]_2}$ 
				uniformly at random for $(u,v) \in \ZZ_2^2$, and sends them to Server A.

\item[2)] \textbf{Unitary operation 1}: 
Server A sets the initial state $|0\rangle^{\otimes n}$ on ${\cal H}^{\otimes n}$.
Server A applies unitaries
$\sX_s^u \sT_s(Q_{2,1,u,s} \by_{1,s}) \sX_s^u$ to ${\cal H}_s$
and $\sX_s^u \sX_t^v
 \sCZ_{(s,t)}(Q_{3,1,(u,v),(s,t)} \bz_{1,(s,t)}) \sX_s^u \sX_t^v$ 
to ${\cal H}_{s}\otimes {\cal H}_{t}$.
Server A applies the Bell measurement 
$\{|\Phi_{a,b}\rangle\}_{a,b \in \ZZ_2}$
on ${\cal H}_s\otimes {\cal H}_{A,1,s}$ for $s=1, \ldots, n$.
Then, Server A obtains the outcome $(A_{1,s},B_{1,s})$ and sends them to User 1
for $s=1, \ldots, n$.
Server A applies unitary
$\sZ_s^{B_{1,s}} \sX_s^{A_{1,s}}$ to ${\cal H}_{A,2,s}$.

\item[3)] \textbf{Query 2}: 
User 1 generates
				$\bQ_{2,1,u}'=(Q_{2,1,u,s}')_{s \in [n]}  \in  \ZZ_8^{n}$ 
				and
				$\bQ_{3,1,(u,v)}'=(Q_{3,1,(u,v),(s,t)}')_{(s,t) \in [n]_2}
				\in  \ZZ_2^{[n]_2}$ as
\begin{align}
Q_{2,1,u,s}' &:=-Q_{2,1,u- A_{1,s},s}
+y_{1,s}'\delta_{u,A_{1,s}} ,\Label{MU1}
\end{align}
and
\begin{align}
&Q_{3,1,(u,v),(s,t)}' \nonumber \\
&:=-Q_{3,1,(u- A_{1,s},v- A_{1,t}),(s,t)}
+z_{1,(s,t)}'\delta_{u,A_{1,s}} \delta_{v,A_{0,t}+A_{1,t}} .
\Label{MU2}
\end{align}
User 1 generates
				$\bQ_{1,1,u}'=(Q_{1,1,u,s}')_{s \in [n]}  \in  \ZZ_4^{n}$ 
according to the uniform distribution. 
User 1 sends them to Server B.

\item[4)] \textbf{Unitary operation 2}: 
(i) Server B applies unitaries
$\sX_s^u \sT_s(Q_{2,1,u,s}' \by_{1,s}) \sX_s^u$ to ${\cal H}_{B,1,s}$
and $\sX_s^u \sX_t^v
 \sCZ_{(s,t)}(Q_{3,1,(u,v),(s,t)}' \bz_{1,(s,t)}) \sX_s^u \sX_t^v$ 
to ${\cal H}_{B,1,s}\otimes {\cal H}_{B,1,t}$ for $(u,v) \in \ZZ_2^2$.
(ii) Then,
Server B applies unitaries
$\sX_s^u \sH_s(Q_{1,1,u,s}' \by_{1,s}) \sX_s^u$ to ${\cal H}_{B,1,s}$
for $u \in \ZZ_2$.
(iii) Server B applies the Bell measurement 
$\{|\Phi_{a,b}\rangle\}_{a,b \in \ZZ_2}$
on ${\cal H}_{B,1,s}\otimes {\cal H}_{B,2,s}$ for $s=1, \ldots, n$.
Then, Server B obtains the outcome $(A_{2,s},B_{2,s})$ and sends them to Users 1 and 2
for $s=1, \ldots, n$.
(iv) Server B applies unitary
$\sZ_s^{B_{2,s}} \sX_s^{A_{2,s}}$ to ${\cal H}_{B,3,s}$.
\end{description}
\noindent {\bf We perform the following steps for $j=2,\ldots,m $.}
\begin{description}
\item[4$j$-3)] \textbf{Query 2$j$-1}: 
User $j-1$ generates
				$\bQ_{1,j-1,u}=(Q_{1,j-1,u,s})_{s \in [n]}  \in  \ZZ_4^{n}$ as
\begin{align}
&Q_{1,j-1,u,s} \nonumber \\
:=&-Q_{1,j-1,u- (A_{2j-2,s}+B_{2j-2,s}+A_{2j-3,s}+B_{2j-3,s}),s}' \nonumber \\
&+x_{j-1,s}'\delta_{u,
A_{0,s}+B_{0,s}+A_{2j-2,s}+B_{2j-2,s}} \Label{ZMTX}
\end{align}
		Then, User $j-1$ sends them to Server A. 
Also, User $j$ generates
				$\bQ_{2,j,u}=(Q_{2,j,u,s})_{s \in [n]}  \in  \ZZ_8^{n}$ 
				and
				$\bQ_{3,j,(u,v)}=(Q_{3,j,(u,v),(s,t)})_{(s,t) \in [n]_2}
				\in  \ZZ_2^{[n]_2}$ 
				uniformly at random for $(u,v) \in \ZZ_2^2$.
		Then, User $j$ sends them to Server A. 

\item[4$j$-2)] \textbf{Unitary operation  2$j$-1}: 
Server A applies unitaries
$\sX_s^u \sH_s(Q_{1,j-1,u,s} \by_{j-1,s}) \sX_s^u$ to ${\cal H}_{A,2j-2,s}$
for $u \in \ZZ_2$.
Then, 
Server A applies unitaries
$\sX_s^u \sT_s(Q_{2,j,u,s} \by_{j,s}) \sX_s^u$ to ${\cal H}_{A,2j-2,s}$
and $\sX_s^u \sX_t^v
 \sCZ_{(s,t)}(Q_{3,j,(u,v),(s,t)} \bz_{j,(s,t)}) \sX_s^u \sX_t^v$ 
to ${\cal H}_{A,2j-2,s}\otimes {\cal H}_{A,2j-2,t}$.
Server A applies the Bell measurement 
$\{|\Phi_{a,b}\rangle\}_{a,b \in \ZZ_2}$
on ${\cal H}_{A,2j-2,s}\otimes {\cal H}_{A,2j-1,s}$ for $s=1, \ldots, n$.
Then, Server A obtains the outcome $(A_{2j-1,s},B_{2j-1,s})$ and sends them to 
User $j$ for $s=1, \ldots, n$.
Server A applies unitary
$\sZ_s^{B_{2j-1,s}} \sX_s^{A_{2j-1,s}}$ to ${\cal H}_{A,2j,s}$.

\item[4$j$-1)] \textbf{Query 2$j$}: 
User $j$ generates
				$\bQ_{2,j,u}'=(Q_{2,j,u,s}')_{s \in [n]}  \in  \ZZ_8^{n}$ 
				and
				$\bQ_{3,j,(u,v)}'=(Q_{3,j,(u,v),(s,t)}')_{(s,t) \in [n]_2}
				\in  \ZZ_2^{[n]_2}$ as
\begin{align}
Q_{2,j,u,s}' &:=-Q_{2,j,u- A_{2j-1,s}-A_{2j-2,s},s}
+y_{j,s}'\delta_{u,A_{0,s}+A_{2j-1,s}} 
\end{align}
and
\begin{align}
&Q_{3,j,(u,v),(s,t)}' \nonumber \\
:=&-Q_{3,j,(u- A_{2j-1,s}-A_{2j-2,s},v- A_{2j-1,t}-A_{2j-2,t}),(s,t)}\nonumber \\
&+z_{j,(s,t)}' \delta_{u,A_{0,s}+A_{2j-1,s}} 
\delta_{v,A_{0,t}+A_{2j-1,t}} .
\end{align}
User $j$ generates
				$\bQ_{1,j,u}'=(Q_{1,j,u,s}')_{s \in [n]}  \in  \ZZ_4^{n}$ 
according to the uniform distribution. 
User $j$ sends them to Server B.

\item[4$j$)] \textbf{Unitary operation 2j}: 
Server B applies unitaries
$\sX_s^u \sT_s(Q_{2,j,u,s}' \by_{j1,s}) \sX_s^u$ to ${\cal H}_{B,2j-1,s}$
and $\sX_s^u \sX_t^v
 \sCZ_{(s,t)}(Q_{3,j,(u,v),(s,t)}' \bz_{j,(s,t)}) \sX_s^u \sX_t^v$ 
to ${\cal H}_{B,2j-1,s}\otimes {\cal H}_{B,2j-1,t}$ for $(u,v) \in \ZZ_2^2$.
Then,
Server B applies unitaries
$\sX_s^u \sH_s(Q_{1,j,u,s}' \by_{j,s}) \sX_s^u$ to ${\cal H}_{B,2j-1,s}$
for $u \in \ZZ_2$.

Server B applies the Bell measurement 
$\{|\Phi_{a,b}\rangle\}_{a,b \in \ZZ_2}$
on ${\cal H}_{B,2j-1,s}\otimes {\cal H}_{B,2j,s}$ for $s=1, \ldots, n$.
Then, Server B obtains the outcome $(A_{2j,s},B_{2j,s})$ and sends them to 
Users $j$ and $j+1$ for $s=1, \ldots, n$.
Server B applies unitary
$\sZ_s^{B_{2j,s}} \sX_s^{A_{2j,s}}$ to ${\cal H}_{B,2j+1,s}$.
\end{description}

\noindent {\bf After Steps with $j=m$, we perform the following remaining steps.}

\begin{description}
\item[4$m$+1)] \textbf{Query 2$m$+1}: 
User m generates
				$\bQ_{1,m,u}'=(Q_{1,m,u,s}')_{s \in [n]}  \in  \ZZ_4^{n}$ as \eqref{ZMTX}
				with $j=m+1$.
		Then, User m sends them to Server A. 

\item[4$m$+2)] \textbf{Unitary operation  2m+1}: 
(i) Server A applies unitaries
$\sX_s^u \sH_s(Q_{1,m,u,s} \by_{m,s}) \sX_s^u$ to ${\cal H}_{A,2m,s}$
for $u \in \ZZ_2$.
(ii) Then, Server A 
applies the measurement $\{ |0\rangle, |1\rangle \}$
to the system ${\cal H}_{A,2m,s}$ and obtains the outcome
$X_s $
for $s=1, \ldots, n_{\circ}$.
(iii) Server A sends
the outcome $X_s $
for $s=1, \ldots, n_{\circ}$ to User $m+1$.

\item[4$m$+3)] \textbf{Construction}:
The user calculates 
$A_{2m,s}+X_{s}$ for $s=1, \ldots, n_{\circ}$. 
\end{description}
\end{prot}

\begin{theo}
Protocol \ref{Pro-C} is a correct TGDMQC protocol that satisfies
the user-secrecy even with dishonest servers and 
the server-secrecy even with dishonest users.
Its upload complexity is $(4n^2+16n)m$ bits.
Its download complexity is $4nm+n_{\circ}$ bits.
\end{theo}

Since Protocol \ref{Pro-C} works as a TGDMQC protocol,
we can realize a variant of TOQC without quantum communication, as
explained in Section \ref{S434}.

\begin{proof}
\noindent{\bf Complexity}: 
The calculation of complexity of Protocol \ref{Pro-C} 
is quite similar to 
the calculation of complexity of Protocol \ref{Pmain}. 
Their difference is the following.
While 
Step 1) of Protocol \ref{Pmain} has transmission of a quantum system with $n$ qubits,
Step 1) of Protocol \ref{Pro-C} has no transmission of quantum system.
While Step 1) of Protocol \ref{Pmain} has transmission of a quantum system with 
$n_{\circ}$ qubits,
Step 1) of Protocol \ref{Pro-C} has transmission of 
$n_{\circ}$ bits.
Considering this difference, we can calculate the 
the complexity of Protocol \ref{Pro-C} from 
the complexity of Protocol \ref{Pmain}.

\noindent{\bf Correctness}: 

\begin{widetext}
Similar to \eqref{MVA} in the proof of Theorem \ref{Th2},
the state at the end of (i) of Step 3) on $\otimes_{s=1}^n {\cal H}_{B,1,s}$
is the following.
\begin{align}
&|\psi_1\rangle\nonumber \\
&:=\Big(\bigotimes_{(s,t) \in [n]_2}
\prod_{(u,v) \in \ZZ_2^2}
\sX_s^u \sX_t^v
 \sCZ_{(s,t)}(Q_{3,1,(u,v),(s,t)}' \bz_{1,(s,t)}) \sX_s^u \sX_t^v
\Big)
\Big(\bigotimes_{s=1}^n
\prod_{u \in \ZZ_2}
\sX_s^u \sT_s(Q_{2,1,u,s}' \by_{1,s}) \sX_s^u\Big)
\Big(\bigotimes_{s=1}^n \sZ_s^{B_{1,s}} \sX_s^{A_{1,s}} \Big)\nonumber \\
&\quad\cdot\Big(\bigotimes_{(s,t) \in [n]_2}
\prod_{(u,v) \in \ZZ_2^2}
\sX_s^u \sX_t^v
 \sCZ_{(s,t)}(Q_{3,1,(u,v),(s,t)} \bz_{1,(s,t)}) \sX_s^u \sX_t^v
\Big)
\Big(\bigotimes_{s=1}^n
\prod_{u \in \ZZ_2}
\sX_s^u \sT_s(Q_{2,1,u,s} \by_{1,s}) \sX_s^u\Big)
\Big(\bigotimes_{s=1}^n \sZ_s^{B_{0,s}} \sX_s^{A_{0,s}} \Big)
|\psi\rangle\nonumber \\
&\doteq
\Big(\bigotimes_{s=1}^n \sZ_s^{B_{1,s}} \sX_s^{A_{1,s}} \Big)
\Big(\bigotimes_{s=1}^n \sZ_s^{B_{0,s}} \sX_s^{A_{0,s}} \Big)
\Big(\bigotimes_{s \in [n]}\sT_s(\by_{1,s}\cdot \by_{1,s}')\Big)
\Big(\bigotimes_{(s,t) \in [n]_2} \sCZ_{(s,t)}( \bz_{1,(s,t)}\cdot \bz_{1,(s,t)}')\Big)
|0\rangle^{\otimes n},\Label{MVA2}
\end{align}
where the above equation follows from 
\eqref{MU1} and \eqref{MU2}. 

Repeating the same discussion as the proof of Theorem \ref{Th2}, 
we find that 
the state at the end of (i) of Step 4m+2) on $\otimes_{s=1}^n {\cal H}_{A,2m,s}$
is the following.
\begin{align}
|\psi_{2m}\rangle
:=
\Big(\bigotimes_{s=1}^n \sZ_s^{B_{2m,s}} \sX_s^{A_{2m,s}} \Big)
\Bigg(\prod_{j=1}^m \Big(\bigotimes_{s \in [n]}\sH_s(\bx_{j,s}\cdot\bx_{j,s}')\Big)
\Big(\bigotimes_{s \in [n]}\sT_s(\by_{j,s}\cdot\by_{j,s}')\Big)
\Big(\bigotimes_{(s,t) \in [n]_2} \sCZ_{(s,t)}( \bz_{j,(s,t)}\cdot\bz_{j,(s,t)}')\Big) 
\Bigg) |0\rangle^{\otimes n} .
\end{align}
Then, Step 4m+3) constructs the desired outcome.
\end{widetext}

\noindent{\bf Server secrecy}: 
Next, we show the server-secrecy even with dishonest users.
Assume that both servers are honest and 
all users collude and are dishonest.
The users obtain the variables $(A_{k,s},B_{k,s})$ for $k=1, \ldots, 2m$ and $s=1,\ldots,n$,
and the variables $\bX=(X_{1}, \ldots, X_{n_{\circ}})$.
These variables $(A_{k,s},B_{k,s})$
are independent of the program $\bw$.
Hence, 
only the variables $\bX$
are related to  the program $\bw$.
Since its size is the same as the desired output information, 
the users do not obtain any information for 
the program $\bw$ more than the desired output information.
In other words,
the users can generate the final state on their own whole system
by using the desired output information
and classical information generated by themselves.

\noindent{\bf User secrecy}: 
Assume that  a set of users $\Theta \subset [m]$ are honest.
Since $\bQ_{2,j,u}$, $\bQ_{3,j,(u,v)}$, and $\bQ_{1,j,u}$
are independently subject to the uniform distribution for $j \in \Theta$,
Server A obtains no information for $\{(\bx_{j}',\by_{j}',\bz_{j}')\}_{j \in\Theta}$.
In the same way,
Server B obtains no information for $\{(\bx_{j}',\by_{j}',\bz_{j}')\}_{j \in\Theta}$.
We assume that the servers are honest additionally.
User $j'$ with $j'=1,\ldots, m$ receives 
the variables $(A_{2j'-1,s},B_{2j'-1,s})$ and $(A_{2j',s},B_{2j',s})$ for 
$s=1,\ldots,n$,
which are 
independently subject to the uniform distribution.
User $j'$ obtains no information for 
$\{(\bx_{j}',\by_{j}',\bz_{j}')\}_{j \in\Theta\setminus \{j'\}}$ for $j'=1, \ldots, m$.
Also, User $m+1$ obtains no information expect for $\bX$, i.e.,
User $m+1$ can recover
the final state on his/her whole system from $\bX$
and classical information generated by himself/herself.
In addition, even when users in $[m+1]\setminus \Theta$ collude,
they obtain no information except for $\bX$, i.e.,
they can recover
the final state on their whole system from $\bX$
and classical information generated by themselves.
Hence, the user-secrecy holds.
\end{proof}

\section{Conclusion}\Label{S9}
We have proposed a new concept, oblivious quantum computation (OQC),
and have introduced its efficient protocol with two servers.
In this protocol, two servers cannot be communicated with each other,
but share many prior entangled states. 
The input state is masked by the random application of Pauli operators.
The quantum state is transferred between two servers alternately via
quantum teleportation.
The user asks each server to apply various unitaries such that 
unnecessarily unitary operations are canceled
and the request to each server behaves as a completely random request.

Our protocol is based on the universal gate set composed of 
the controlled Z operation, the modification Hadamard gate $\sH$, and the $1/8$-phase gate $\sT$.
When the number of gates is $m$ and the inputs are composed of $n$ qubits,
the communication complexity of our protocol is upper bounded by 
$2n^2 m+ 20 n m$ bits plus $2n$ qubits.
In contrast, 
even when the input is restricted into basis states in the $n$ qubits,
the application of the conventional protocol for two-server oblivious transfer requires
communication complexity with exponential size for $n$.
Hence, when the number $m$ of gates is a polynomial, 
our protocol offers an exponential improvement over existing methods.

\if0
The key point of this paper is composed of two ideas.
The first idea is summarized as Protocol \ref{toy}.
A similar idea to Protocol \ref{toy} was used in the papers \cite{SH21,SH22}.
While Protocol \ref{toy} handles only discrete parameters in $\ZZ_8$,
the protocol in the papers \cite{SH21,SH22} 
handles continuous parameters.
Hence, Protocol \ref{toy} does not need a complicated procedure that appears in 
the protocol in the papers \cite{SH21,SH22} 
related to the continuity of the parameters.

The second idea is summarized as Protocol \ref{Psub}.
Protocol \ref{Psub} converts the Schmidt basis of a given entangled state
by using LOCC.
This idea has not been used until this paper.
Since these two ideas are very useful,
it is expected to apply these two ideas to other topics in quantum information.
That is, it is an interesting future study to find another interesting application of these two ideas.
\fi

In addition, we have introduced an additional concept, 
generalized delegated multiparty quantum computation (GDMQC),
by generalizing delegated multiparty quantum computation (DMQC).
In Section \ref{S8-2}, modifying our two-server protocol for OQC,
we have proposed a two-server protocol for GDMQC,
whose basic idea is the same as our protocol for OQC.

There are still several other remaining problems.
Since the concept, GDMQC, is very broad,
we can expect that it covers various problem settings.
Therefore, it is an interesting future problem to find its fruitful applications.
In addition, the papers \cite{SH19-2,SH20,AHPH20,ASHPHH21,SJ18,FHGHK17}
discussed the case when servers potentially collude, but there are more than two servers.
It is an interesting remaining problem
to extend our results to the above case with colluding servers.

Further, our two-server protocol for GDMQC contains 
a two-server protocol for conventional delegated quantum computation \cite{Childs,BFK,BKBF,MF,Morimae,MDF,MF2,LCWW,SZ,HM}
as a special case by assuming that there is only one user and 
the servers' program $\bw$ is $\be$.
Indeed, it could be possible to verify this delegated quantum computation protocol by inserting check bits.
However, it is not easy to evaluate how many check bits
are needed to achieve a given precision level
because it is difficult to evaluate 
the probability of detecting an error with a given dishonest server's operation.
In fact, once the above detecting probability is evaluated,
it is possible to evaluate the quality of verification
by using existing results of the verification with the non-iid setting \cite{Significance}.
This evaluation is another interesting remaining problem.

\section*{Acknowledgement}
MH was supported in part by the National
Natural Science Foundation of China (Grants No. 62171212).

\end{document}